\documentclass[preprint,tightenlines,showpacs,showkeys,floatfix,nofootinbib,
superscriptaddress,fleqn]{revtex4}
\usepackage{graphicx}
\usepackage{bm}
\usepackage{dcolumn}
\usepackage{amsmath}
\usepackage{amsfonts}
\usepackage{amssymb}
\usepackage{color}

\begin{document}
\preprint{INHA-NTG-05/2008}
\title{Vector transition form factors of the $N K^*\to\Theta^+ $ and
$N \bar{K}^*\to \Sigma_{\overline{10}}^{*-}$ in the SU(3)
chiral quark-soliton model}
\author{Tim Ledwig}
\email{Tim.Ledwig@tp2.rub.de}
\affiliation{Institut f\"ur Theoretische Physik II, Ruhr-Universit\"
at Bochum, D--44780 Bochum, Germany}
\author{Hyun-Chul Kim}
\email{hchkim@inha.ac.kr}
\affiliation{Department of Physics, Inha University, Incheon 402-751,
Republic of Korea}
\author{Klaus Goeke}
\email{Klaus.Goeke@tp2.rub.de}
\affiliation{Institut f\"ur Theoretische Physik II, Ruhr-Universit\"
at Bochum, D--44780 Bochum, Germany}
\date{March 2008}

\begin{abstract}
We investigate the vector transition form factors of the nucleon and
vector meson $K^*$ to the pentaquark baryon $\Theta^+$ within the
framework of the SU(3) chiral quark-soliton model.  We take into
account the rotational $1/N_c$ and linear $m_{\rm s}$ corrections,
assuming isospin symmetry and employing the symmetry-conserving
quantization.  It turns out that the leading-order contributions to
the form factors are almost cancelled by the rotational corrections.
Because of this, the flavor SU(3) symmetry-breaking terms yield
sizeable effects on the vector transition form factors.  In
particular,  the main contribution to the electric-like transition
form factor comes from the wave-function corrections, which is a
consequence of the generalized  Ademollo-Gatto theorem derived in the
present work.  We estimate with the help of the vector meson dominance
the $K^*$ vector and tensor coupling constants for the $\Theta^+$:
$g_{K^{*}N\Theta}=0.74 - 0.87$ and $f_{K^{*}N\Theta}=0.53 - 1.16$.  We
argue that the outcome of the present work is
consistent with the null results of the CLAS experiments in the
reactions $\gamma n\to K^-\Theta^+$ and $\gamma p \to \bar{K}^0
\Theta^+$.  The results of the present work are also consistent with
the recent experiments at KEK.  In addition, we present the results of
the $\Sigma_{\overline{10}}\to N\bar{K}^*$ transition form factors and 
its $\bar{K}^*N\Sigma_{\overline{10}}$ coupling constants.  
\end{abstract}
\maketitle

\section{Introduction}
Since Diakonov, Petrov and Polyakov~\cite{Diakonov:1997mm}
predicted the mass and width of the exotic pentaquark baryon
$\Theta^+$ (with leading Fock-component $uudd\bar{s}$), the
pentaquark baryons have attracted much attention (see also an
earlier estimate of the mass by Prasza\l owicz in the soliton
approach of the Skyrme model~\cite{Praszalowicz}).  After the LEPS 
collaboration has reported the evidence of the $\Theta^+$
first~\cite{Nakano:2003qx}, many experiments have announced its
existence~\cite{Barmin:2003vv,CLAS1,CLAS2,Hermes,
Zeus,neutrino,SVD,COSY_TOF}, while the $\Theta^+$ has not been
seen in almost all high-energy
experiments~\cite{BES,BABAR,BELLE,LEP,ALPEH,HERA,SPHINX,HyperC,CDF}
(see Ref.~\cite{Hicks:2005gp} for a review of the experimental
status until 2005). Actually, a series of very recent CLAS
experiments, i.e. dedicated experiments to search for the
$\Theta^+$, has reported null results of the
$\Theta^+$~\cite{Battaglieri:2005er,McKinnon:2006zv,
Niccolai:2006td,DeVita:2006ha} and has casted doubts on its
existence. On the other hand, the DIANA collaboration has
continued to search for the $\Theta^{+}$ and reported very
recently the formation of a narrow $pK^{0}$ peak with mass
of $1537\pm2$ MeV/$c^{2}$ and width of $\Gamma=0.36\pm0.11$ MeV in
the $K^{+}n\rightarrow K^{0}p$ reaction~\cite{Barmin:2006we}.
Moreover, several other new experiments with positive results for
the $\Theta^{+}$  have been
reported 
~\cite{new_SVD,Hotta:2005rh,Miwa:2006if,Nakano,SVD2:2008}.  Thus, 
it is too early to conclude the absence of the $\Theta^+$ and more
efforts should be put for understanding the $\Theta^+$
theoretically as well as experimentally.  It was also shown
that results of different current experiments may be reconciled with
each other by the specific hypothesis about the $\Theta^+$ production
mechanism \cite{Azimov:2007}.  

In addition to the $\Theta^+$ search, a recent GRAAL
experiment~\cite{Kuznetsov:2004gy,Kuznetsov:2006de,Kuznetsov:2006kt}
has reported results on the cross section of
$\eta$-photoproduction off the deuteron. The authors have
identified a resonant structure around 1.67 GeV in the neutron
channel, while they have not seen it in the quasi-free proton channel.
In the same paper it has been shown that a resonance of 10 MeV width
would be enhanced to the measured 40 MeV width by Fermi motion of the
neutron in the deuteron. This new resonance is consistent with the
theoretical predictions by Ref.~\cite{Diakonov:2003jj,Arndt:2003ga} of
a new nucleon-like state in that mass region. Actually, the narrow
width and its dependence on the initial isospin state are benchmarks
for the photo-excitation of the nonstrange anti-decuplet pentaquark
as was suggested in Ref.~\cite{Polyakov:2003dx}. The cross section of 
photo-excitation of the anti-decuplet proton state should be
suppressed relative to the neutron
one~\cite{Polyakov:2003dx,Kim:2005gz} and in fact it is.  In
the photon beam asymmetry $\Sigma$, however, the corresponding signal
can be enhanced due to interference effects. A new analysis of the
free proton GRAAL data in
Refs.~\cite{Kuznetsov:2007dy,Kuznetsov:2008gm} has revealed the
resonance structure in $\Sigma$ at a mass around 1685 MeV 
with a width of $\leq 15$ MeV. The results of the analysis
of~\cite{Kuznetsov:2007dy} do not agree with those of
Ref.~\cite{Bartalini:2007fg}. A discussion of this point can be
found in Ref.~\cite{Kuznetsov:2008gm}.  Furthermore, the
LNS-GeV-$\gamma$ collaboration~\cite{tohoku,tohoku2} has reported a
new resonance at $1670$ MeV  with a width of $\leq 50$ MeV in the
$\gamma d\to \eta pn$ reaction. This resonance is enhanced in the
$\gamma n\to \eta n$ reaction and in the cross section no coupling
to the quasi-free proton channel is observed.  Moreover, the
CB-ELSA collaboration~\cite{CBELSA} has also announced the data
compatible with those of GRAAL and LNS-GeV-$\gamma$, which are studied 
theoretically in Ref~\cite{Fix:2007st}.  All these experimental
facts are consistent with the results for the transition magnetic
moments in the chiral quark-soliton model
($\chi$QSM)~\cite{Polyakov:2003dx,Kim:2005gz} as well as  
with phenomenological results~\cite{Azimov:2005jj}.  Moreover, recent
theoretical calculations of the $\gamma N\to \eta N$
reaction~\cite{Choi:2005ki,Choi:2007gy} describe qualitatively 
well the GRAAL data, based on the values of the magnetic
transition moments in Refs.~\cite{Kim:2005gz,Azimov:2005jj}.
From these calculations~\cite{Fix:2007st,Choi:2005ki,Choi:2007gy}, one
may consider them as a hint that the $N^*$ resonance seen in the GRAAL
experiment could perhaps be identified as one of the nonstrange
pentaquark baryons.

The null results of the CLAS experiments imply that the total
cross sections of the relevant reactions should be very small.  In
fact, Ref.~\cite{Battaglieri:2005er} found the total cross section
for the $\Theta^+$ in the $\gamma p\to \bar{K}^0 K^+ n$ reaction
to be 0.8 nb in the $95\,\%$ CL upper limit.  Similarly, it was
found to be 0.7 nb in the combined $\gamma p\to \bar{K}^0 K^+ n$
and $\gamma p\to \bar{K}^0 K^0 p$ reactions~\cite{DeVita:2006ha}.
The upper limit of the cross section for the elementary $\gamma n
\to K^-\Theta^+ $ process was estimated to be around 3
nb~\cite{McKinnon:2006zv}.  In fact, these small numbers are
consistent with predictions of Ref.~\cite{Azimov:2006he} and of
Ref.~\cite{Kwee:2005dz} prior to the CLAS measurements.  It is known
from theoretical works~\cite{Kwee:2005dz,Oh:2003kw,Nam:2004xt} that 
vector meson $K^*$ exchange plays an essential role in describing
the mechanism of both the $\gamma n\to K^- \Theta^+$ and $\gamma
p\to \bar{K}^0 \Theta^+$ reactions, the parity of the $\Theta^+$
being assumed to be positive. In particular, the total cross
section of the $\gamma p\to \bar{K}^0 \Theta^+$ reaction is rather
sensitive to the contribution of $K^*$ exchange, since other
contributions turn out to be very tiny.  Since, however, there is
no solid information on the coupling strength for the $NK^*
\to\Theta^+$ vertex theoretically as well as experimentally, it is
crucial to provide some theoretical guideline to estimate the
coupling strength for the $K^*N\Theta^+$ vertex.

Thus, in the present work, we aim at investigating the vector
transition form factors for the $N K^*\to\Theta^+$.  Since the
mass difference between the $\Theta^+$ and $N$ is not at all
small, the divergence of the transition matrix elements does not
vanish. While the vector transition form factors can be
interpreted as the $K^*$ coupling strengths, the divergence of the
transition vector current is related to the coupling strength for
the scalar meson $\kappa$.  However, we want to concentrate in
this work on the vector transition form factors for the
$NK^*\to\Theta^+$.  It can be shown that the $NK^* \to \Theta^+$
transition form factor of the time component of the vector
current, i.e. the electric-like transition form factor is
suppressed due to a generalization of the Ademollo-Gatto theorem.
The transition form factor of the time component of the vector
current at $Q^2=0$ arises only from the wave-function corrections,
i.e. from the mixing angle between the octet and anti-decuplet
representations. Indeed, we will show in the present work that the
electric-like transition form factor for the $N K^*\to\Theta^+$ is
suppressed, based on the $\chi$QSM with isospin symmetry and
symmetry-conserving quantization~\cite{Praszalowicz:1998jm}
imposed.  For completeness, we will also present the results for
the vector transition form factors and coupling constants for the
$N \bar{K}^*\to \Sigma_{\overline{10}}^{-}$.

The $\chi$QSM has been proved very successful not only in
predicting the $\Theta^+$ but also even more in describing various
properties of SU(3) baryon octet and decuplet such as the mass
splittings, form factors and parton and
antiparton distributions~\cite{Meissner:1989kg,Goeke:1991fk,
Alkofer:1994ph,Christov:1995vm,Dressler:2001, 
GoekePPSU:2001,SchweitzerUPWPG:2001,OssmannPSUG:2005,KimPPYG:2005,
SilvaKUG:2005,WakamatsuN:2006,Wakamatsu:2005,Wakamatsuref:2003,
Wakamatsu:2000}.  In particular, the dependence of almost all form
factors on the momentum transfer is well reproduced within the
$\chi$QSM.  As a result, the strange electromagnetic form
factors~\cite{Goeke:2006gi} and the parity-violating asymmetries
of polarized electron-proton scattering, which require nine
different form factors (six electromagnetic form factors
$G_{E,M}^{(u,d,s)} (Q^2)$ and three axial-vector form factors
$G_{A}^{(u,d,s)} (Q^2)$), are in good agreement with experimental
data~\cite{Silva:2005qm}.  Therefore, it is worthwhile to extend
the study of the form factors to the baryon anti-decuplet within
the $\chi$QSM. Such a study is particularly interesting since in
the $\chi$QSM the $d\overline{s}$-component of e.g. the
$\Theta^{+}$ does not consist of valence quarks (as e.g. all the
quark models suggest), but is formed by the collective excitation
of the chiral mean field generated by the rotation of the soliton.
In the case of the $\Theta^{+}$ ($uud(d\overline{s}$)) this collective 
excitation carries the quantum numbers of a valence
$d\overline{s}$-pair. Thus the leading Fock-component of the
$\Theta^{+}$ is $uudd\overline{s}$, however, probably with many
subleading terms.

We sketch the present work as follows. In Section II, we
describe briefly the general formalism about the vector
$NK^*\to\Theta^+$ transition form factors.  In Section III, we
explain the vector-meson dominance to relate the electromagnetic-like
transition form factors to the vector and tensor coupling constants
for the $K^*N\Theta^+$ vertex.  In Section IV, we generalize the
Ademollo-Gatto theorem in the context of the $N\to\Theta^+$
transition.  Section V shows how to derive the transition form factors
in the $\chi$QSM.  In Section VI and Section VII, we present our
results for the transition form factors and coupling
constants and and discuss them.  In the final Section we summarize the
present work and draw conclusions.  Some detailed expressions can be
found in the Appendix.
\section{General Formalism}
We start with the vector current relevant for the $NK^*\to\Theta^+$
transition.  We will concentrate here only on the $nK^*\to\Theta^{+}$
transition, since the $pK^0 \to\Theta^{+}$ one can be directly
obtained due to the isospin relation that yields an overall
factor $-1$.  The relevant vector current for this process is defined as
\begin{equation}
J_{V}^{\mu}(x)=\overline{\psi}(x)\gamma^{\mu}\frac{1}{2}
\Big(\lambda^{4}-i\lambda^{5}\Big)\psi(x)=\overline{s}(x)\gamma^{\mu}u(x),
\label{eq:rel_current}
\end{equation}
where $\psi(x)=(u(x),d(x),s(x))$ denote the corresponding
quark fields and $\lambda^{a}$ are flavor $SU(3)$ Gell-Mann matrices.
The matrix element of the vector current in Eq.(\ref{eq:rel_current})
can be expressed in terms of three real transition form factors
$F_i^{n\Theta}$ as follows:
\begin{equation}
 \langle\Theta^+(p^{\prime})| \overline{s}(0) \gamma^{\mu}u(0)|
 n(p)\rangle=\overline{u}_{\Theta}(\bm p^{\prime}, s')
\left[F_{1}^{n\Theta}(q^{2}) \gamma^{\mu}+\frac{F_{2}^{n\Theta}(q^{2})i
\sigma^{\mu\nu}q_{\nu}}{M_{\Theta}+M_{n}}+\frac{F_{3}^{n\Theta}(q^{2})q^{\mu}}{
M_{\Theta}+M_{n}}\right]u_{n}(\bm p, s),
\label{general}
\end{equation}
where $q^{2}=-Q^2$ is the square of the four momentum transfer
$q=p'-p$.  The $u_{n(\Theta)}$ represents the Dirac spinor
for the neutron ($\Theta^+$) with momentum $\bm p$ ($\bm p'$) and spin
$s$ ($s'$).  The $M_{n(\Theta)}$ stands for the mass of the neutron
($\Theta^{+}$).  Because of the mass difference between the neutron
and the $\Theta^+$, the vector current is not conserved in this case:
The matrix element of its divergence is then given as~\footnote{As one
can see, the divergence of the vector transition current is related
to the matrix element of the scalar operator $\overline{s}u$, which
indicates that the combination of the transition form factor $F_{1}$
and $F_{3}$ will provide information on the scalar meson $\kappa$
coupling constant for the $\kappa N\Theta^+$ vertex.}
\begin{equation}
(m_{u}-m_{s})\langle\Theta^+(p')|
\overline{s}u|n(p) \rangle = \overline{u}_{\Theta}(\bm p^{\prime},s')
u_{n}(\bm p, s)\left[F_{1}^{n\Theta}(q^{2})(M_{\Theta}-M_{n})
+ \frac{F_{3}^{n\Theta}(q^{2})q^{2}}{M_{\Theta}+M_{n}}\right].
\label{divergence}
\end{equation}
From now on we will assume isospin symmetry and use the average
$\overline{m}=(m_{u}+m_{d})/2$ for the up- and down-quark
masses.  The $m_{\mathrm{s}}$ denotes
the strange current quark mass.

In the present scheme, it is more convenient to calculate the
Sachs-type transition form factors $G^{n\Theta}_{E}$ and
$G^{n\Theta}_{M}$, which we will now denote from now on as electric-like
and magnetic-like transition form factors, since they are directly related
to the matrix elements of the time and space components of the vector
transition current~\cite{Kim_eleff,ChristovEMFF}:
\begin{eqnarray}
G_{E}^{n\Theta}(Q^2) &=& \int\frac{d\Omega_{q}}{4\pi}\langle\Theta(p^{\prime})
|J_{V}^{0}(0)|n(p)\rangle,\cr
G_{M}^{n\Theta}(Q^2) &=& 3M_{n} \int \frac{d\Omega_{q}}{4\pi} \frac{q^{i}
  \epsilon^{ik3}}{i\mid \bm q\mid^{2}} \langle\Theta(p^{\prime})|
J_{V}^{k}(0)|n(p)\rangle.
\label{GM}
\end{eqnarray}
Choosing the rest frame of the $\Theta^{+}$, i.e. $p^{\prime}=(M_{\Theta},\,0)$
and $p=(E_{n},\,-{\bm q})$ and using Eq.(\ref{GM}), we obtain the
following relations:
\begin{eqnarray}
G_{E}^{n\Theta}(Q^{2}) & = & \sqrt{\frac{E_{n}+M_{n}}{2M_{n}}}\left[
F_{1}^{n\Theta}(Q^{2}) - \frac{F_{2}^{n\Theta}(Q^{2})}{M_{\Theta}+M_{n}} \frac{{\bm
    q}^{2}}{E_{n}+M_{n}}+F_{3}^{n\Theta}(Q^{2})
\frac{q^{0}}{M_{\Theta}+M_{n}}\right],\cr
G_{M}^{n\Theta} (Q^{2}) & = & \sqrt{
  \frac{2M_{n}}{E_{n}+M_{n}}}\left[F_{1}^{n\Theta}(Q^{2})+F_{2}^{n\Theta}(Q^{2})
\right].
\label{eq:effGm}
\end{eqnarray}
The kinematics in this frame is given by
\begin{equation}
{\bm q}^{2} = \left(\frac{Q^{2}+M_{\Theta}^{2}+M_{n}^{2}}{
  2M_{\Theta}}\right)^{2}-M_{n}^{2},\;\;\;
 E_{n}=\frac{Q^{2}+M_{n}^{2}+M_{\Theta}^{2}}{2M_{\Theta}}.
\label{vecqdef}
\end{equation}
 In order to separate $F_{i}^{n\Theta}(Q^{2})$ from the Sachs-type form factors,
we need one more form factor, that is, the scalar transition form
factor.  However, since we are interested only in the $K^{*}\Theta N$
coupling constants, we will concentrate in the present work only on the
electromagnetic-like transition form factors $G_E^{n\Theta}(Q^{2})$ and
$G_{M}^{n\Theta}(Q^{2})$.
\section{Vector Meson Dominance\label{sec:vmd}}
In order to determine the coupling constants of the vector meson $K^{*}$
to the nucleon and $\Theta^{+}$, we want to use the vector
meson dominance (VMD)~\cite{Sakurai:1960ju,Feynman:1973xc}.  In the
VMD, the corresponding transition vector current is identified by the
current-field identity (CFI)~\cite{Sakurai:1960ju} as
\begin{equation}
J_{V}^{\mu}(x) = \overline{s}(x)\gamma^{\mu}u(x) =
\frac{m_{K^{*}}^{2}}{f_{K^{*}}} K^{*\mu}(x),
\label{eq:CFI}
\end{equation}
where $m_{K^{*}}=892$ MeV denotes the mass of the vector meson $K^{*}$.
The generalized $K^{*}$ meson coupling constant $f_{K^{*}}$ can
be determined by the following relation~\cite{Riazuddin:1966sw}:
\begin{equation}
f_{K^{*}}^{2} = \frac{m_{K^{*}}^{2}}{m_{\rho}^{2}}f_{\rho}^{2}
\end{equation}
with
\begin{equation}
f_{\rho}^{2} = \frac{\alpha^{2}m_{\rho}4\pi}{3\Gamma(\rho^{0}\to
  e^{+}e^{-})}\,,
\end{equation}
where $m_{\rho}=770$ MeV and $f_{\rho}$ are the $\rho$-meson mass
and the photon-$\rho$ meson coupling constant, respectively. The
$\alpha$ is the electromagnetic fine structure constant and the decay
width $\Gamma(\rho^{0}\to e^{+}e^{-})$ is given by~\cite{Yao:2006px}
as $(7.02\pm0.11)$ keV, from which we get the photo-coupling
constants $f_\rho$ and $f_{K^{*}}$:
\begin{equation}
f_{\rho}\simeq4.96,\;\;\; f_{K^{*}}\simeq5.71.
\end{equation}
Using the CFI in Eq.(\ref{eq:CFI}) and VMD, we can express
Eq.(\ref{general}) as follows:
\begin{eqnarray}
\langle\Theta(p^{\prime})|\overline{s}\gamma^{\mu}u|n(p)\rangle & = & 
\frac{m_{K^{*}}^{2}}{f_{K^{*}}}\,\frac{1}{m_{K^{*}}^{2}-q^{2}}\,\langle
\Theta(p^{\prime})|K^{*\mu}|n(p)\rangle,\cr 
\langle\Theta(p^{\prime})|K^{*\mu}|n(p)\rangle & = & 
\overline{u}_{\Theta} (\bm p^{\prime},s') \left[g_{K^{*}n\Theta}(Q^{2})
\gamma^{\mu}+f_{K^{*}n\Theta}(Q^{2}) \frac{i\sigma^{\mu\nu}q_{\nu}}{
  M_{\Theta}+M_{n}}  \right]u_{n}(\bm p, s),
\label{eq:VDM2}
\end{eqnarray}
where the $g_{K^{*}n\Theta}(Q^{2})$ and $f_{K^{*}n\Theta}(Q^{2})$
denote the vector and tensor coupling constants for the $K^{*}\Theta
n$ vertex, respectively, and they are related to the Diac and
Pauli transition form factors $F_1^{n\Theta}(Q^2)$ and $F_2^{n\Theta}(Q^2)$ in
Eq.(\ref{general}):
\begin{equation}
g_{K^{*}n\Theta}(Q^{2}) =
\frac{f_{K^{*}}(m_{K^{*}}^{2}-q^{2})}{m_{K^{*}}^{2}}\,F_{1}^{n\Theta}(Q^{2}),\;\;
f_{K^{*}n\Theta}(Q^{2}) = \frac{f_{K^{*}}(m_{K^{*}}^{2}-q^{2})}{m_{K^{*}}^{2}}\,
F_{2}^{n\Theta}(Q^{2}).
\label{eq:gf}
\end{equation}
In fact, there is a third coupling constant is related to the $\kappa$
coupling constant together with the vector coupling constant 
$g_{K^{*}n\Theta}$.  However, we will drop it, since it is not
relevant for the present work as discussed before.  We will use 
Eq.(\ref{eq:gf}) later in order to estimate the coupling constants
from the vector transition form factors calculated in the $\chi$QSM.

\section{Form Factors in the $\chi$QSM }
In this Section, we briefly review how to derive the vector
transition form factors for the $n K^{*+}\to\Theta^+$ process within
the framework of the $\chi$QSM.  The details can be found in
Refs.~\cite{Christov:1995vm,Kim_eleff,ChristovEMFF,Silva_Delta}.
The SU(3) $\chi$QSM is characterized by the following partition
function in Euclidean space:
\begin{equation}
\label{eq:part}
\mathcal{Z}_{\mathrm{\chi QSM}} = \int \mathcal{D}\psi\mathcal{D}\psi^\dagger
 \mathcal{D}\pi\exp\left[-\int d^4 x\psi^\dagger D(\pi)\psi\right]
= \int \mathcal{D}\pi\exp(-S_{\mathrm{eff}}[\pi]),
\end{equation}
where $\psi$ and $\pi$ denote the quark and pseudo-Goldstone boson
fields, respectively.  The $S_{\mathrm{eff}}$ stands for the effective
chiral action expressed as
\begin{equation}
  \label{eq:echl}
S_{\mathrm{eff}} = -N_c\mathrm{Tr}\ln D(\pi),
\end{equation}
where $\mathrm{Tr}$ represents the functional trace, $N_c$ the number
of colors, and $D$ the Dirac differential operator in Euclidean space:
\begin{equation}
 \label{eq:Dirac}
D(U) = \gamma_4(i\rlap{/}{\partial} -\hat{m} -MU^{\gamma_5}) = \partial_4
+ h(U) + \delta m.
\end{equation}
The $\hat{m}$ denotes the current quark matrix $\hat{m} =
\mathrm{diag}(\overline{m},\,\overline{m},\,m_{\mathrm{s}})$, isospin
symmetry being assumed.  The $\partial_4$ designates the derivative
with respect to the Euclidean time and $h(U)$ stands for the Dirac
single-quark Hamiltonian:
\begin{equation}
h(U)=-i\gamma_4\gamma_i\partial_i + \gamma_4 MU^{\gamma_5}
+\gamma_4\overline{m}.
\label{eq:diracham}
\end{equation}
The $\delta m$ is the the matrix of the decomposed current quark
masses:
\begin{equation}
  \label{eq:deltam}
\delta m = M_1\gamma_4\bm 1 + M_8 \gamma_4\lambda^8,
\end{equation}
where $M_1$ and $M_8$ are singlet and octet components of the current
quark masses defined as $M_1 =(-\overline{m}+m_{\mathrm{s}})/3$ and
$M_8=(\overline{m}-m_{\mathrm{s}})/\sqrt{3}$.  The $\overline{m}$ is
the average of up- and down-quark masses.  The chiral field
$U^{\gamma_5}$ is written as
\begin{equation}
U^{\gamma_5} = \exp(i\gamma_5 \lambda^a\pi^a) =
\frac{1+\gamma_5}{2}U + \frac{1-\gamma_5}{2} U^\dagger
\end{equation}
with $U=\exp(i\lambda^a\pi^a)$.  We assume here Witten's trivial
embedding of $SU(2)$ into $SU(3)$:
\begin{equation}
  \label{eq:embed}
U_{\mathrm{SU(3)}} = \left(\begin{array}{lr} U_{\mathrm{SU(2)}} & 0
    \\ 0 & 1   \end{array} \right)
\end{equation}
with the SU(2) hedgehog chiral field
\begin{equation}
  \label{eq:hedgehog}
U_{\mathrm{SU2}}=\exp[i\gamma_5 \hat{\bm
  n}\cdot\bm\tau P(r)].
\end{equation}
In order to solve the partition function in
Eq.(\ref{eq:part}), we have to take the large $N_c$ limit and solve it
in the saddle-point approximation, which corresponds at the classical
level to finding the profile function $P(r)$ in
Eq.(\ref{eq:hedgehog}).  In fact, the profile function can be obtained
by solving numerically the functional equation coming from $\delta
S_{\mathrm{eff}}/\delta P(r) =0$, which yields a classical soliton
field $U_c$ constructed from a set of single quark energies $E_n$ and
corresponding states $|n\rangle$ related to the eigenvalue equation
$h(U)|n\rangle =E_n|n\rangle$.

Since the classical soliton does not have the quantum number of the
baryon states, we need to restore them by the semiclassical
quantization of the rotational and translational zero modes.  Note
that the zero modes can be treated exactly within the functional
integral formalism by introducing collective coordinates.  Detailed
formalisms can be found in Refs.~\cite{Christov:1995vm,Kim_eleff}.
Considering the rigid rotations and translations of the classical
soliton $U_c$, we can express the soliton field as
\begin{equation}
U(\bm x, t) = A(t)U_c(\bm x - \bm z(t))A^\dagger (t),
\end{equation}
where $A(t)$ denotes a unitary time-dependent SU(3) collective
orientation matrix and $\bm z(t)$ stands for the time-dependent
displacement of the center of mass of the soliton in coordinate
space.

Having introduced the zero modes as mentioned above, the
Dirac operator in Eq.(\ref{eq:Dirac}) is changed to the following
form:
\begin{equation}
D(U)=T_{z(t)}A(t) \left[ D(U_{c}) + i\Omega(t) -
  \dot{T}_{z(t)}^{\dagger}T_{z(t)}+i\gamma^{4}A^{\dagger}(t)\delta m
  A(t) \right]T_{z(t)}^{\dagger}A^{\dagger}(t),
\end{equation}
where the $T_{z(t)}$ denotes the translational unitary operator and
the $\Omega(t)$ represents the angular velocity of the soliton that is
defined as
\begin{equation}
\Omega=-iA^{\dagger}\dot{A}=-\frac{i}{2}\textrm{Tr} (
A^{\dagger}\dot{A}\lambda^{\alpha})\lambda^{\alpha}=\frac{1}{2}
\Omega_{\alpha}\lambda^{\alpha}.
\end{equation}
Assuming that the soliton rotates and moves slowly, we can treat the
$\Omega(t)$ and $\dot{T}_{z(t)}^{\dagger}T_{z(t)}$ perturbatively.
Moreover, since the flavor SU(3) symmetry is broken weakly, we can
also deal with $\delta m$ perturbatively.

Having quantized collectively, we obtain the following collective
Hamiltonian
\begin{equation}
  \label{eq:Ham}
H_{\textrm{coll}}=H_{\mathrm{sym}} + H_{\mathrm{sb}},
\end{equation}
where $H_{\mathrm{sym}}$ and $H_{\mathrm{sb}}$ represent the SU(3)
symmetric and symmetry-breaking parts, respectively:
\begin{eqnarray}
H_{\mathrm{sym}} &=& M_{c} + \frac{1}{2I_{1}}\sum_{i=1}^3 J_{i}J_{i} +
\frac{1}{2I_{2}} \sum_{a=4}^7 J_{a} J_{a} + \frac{1}{\overline{m}} M_{1}
\Sigma_{SU(2)} ,\cr
H_{\mathrm{sb}} &=& \alpha D_{88}^{(8)}(A) + \beta Y +
\frac{\gamma}{\sqrt{3}} D_{8i}^{(8)}(A)J_{i}.
\end{eqnarray}
The $M_c$ denotes the mass of the classical soliton and $I_{i}$ and
$K_{i}$ are the moments of inertia of the
soliton~\cite{Christov:1995vm}, of which the corresponding
expressions can be found in Ref.~\cite{Blotz:1992pw} explicitly.
The components $J_i$ denote the spin generators and $J_a$ correspond
to those of right rotations in flavor SU(3).  The $\Sigma_{SU(2)}$ is
the $SU(2)$ pion-nucleon sigma term. The $D_{88}^{(8)}(A)$ and
$D_{8i}^{(8)}(A)$ stand for the SU(3) Wigner $D$ functions in the
octet representation.  The $Y$ is the hypercharge operator.  The
parameters $\alpha$, $\beta$, and $\gamma$ in the symmetry-breaking
Hamiltonian are expressed, respectively, as follows:
\begin{equation}
\alpha = \frac{1}{\overline{m}} \frac{1}{\sqrt{3}} M_{8}
\Sigma_{SU(2)} - \frac{N_{c}}{\sqrt{3}} M_{8} \frac{K_{2}}{I_{2}},
\;\;\;\;
\beta = M_{8} \frac{K_{2}}{I_{2}}\sqrt{3},\;\;\;\;
\gamma = -2\sqrt{3}M_{8}\left(\frac{K_{1}}{I_{1}} -
  \frac{K_{2}}{I_{2}}\right).
\end{equation}
The collective wave-functions of the Hamiltonian in Eq.(\ref{eq:Ham})
can be found as SU(3) Wigner $D$ functions in representation
$\mathcal{R}$:
\begin{equation}
  \label{eq:Wigner}
\langle A|\mathcal{R},B(YII_{3},Y^{\prime}JJ_{3}) \rangle =
\Psi_{(\mathcal{R}^{*};Y^{\prime}JJ_{3})}^{(\mathcal{R};YII_{3})}(A) =
\sqrt{\textrm{dim}(\mathcal{R})}\,(-)^{J_{3}+Y^{\prime}/2}\,
D_{(Y,I,I_{3})(-Y^{\prime},J,-J_{3})}^{(\mathcal{R})*}(A).
\end{equation}
The $Y'$ is related to the eighth component of the angular velocity
$\Omega$ that is due to the presence of the discrete valence quark
level in the Dirac-sea spectrum.  Its presence has no effect on the
chiral field, so that it is constrained to be $Y'=-N_c/3=-1$.  In fact,
this constraint allows us to have only the SU(3) representations with
zero triality.

The effects of flavor SU(3) symmetry breaking having been taken into
account, the collective baryon states are not in a pure
representation but start to get mixed with other representations.
This can be treated by considering the second-order perturbation for
the collective Hamiltonian:
\begin{equation}
|B_{\mathcal{R}}\rangle = |B_{\mathcal{R}}^{\mathrm{sym}} \rangle -
\sum_{\mathcal{R}^{\prime}\neq\mathcal{R}} |B_{\mathcal{R}^{\prime}}
\rangle \frac{\langle B_{\mathcal{R}^{\prime}}|\,
  H_{\textrm{sb}}   \,|B_{\mathcal{R}}\rangle}{M(\mathcal{R}^{\prime}) -
  M(\mathcal{R})}\,.
\label{wfc}
\end{equation}
Then, the collective octet and anti-decuplet baryon wave-functions
result in the following expressions~\cite{mixings:2004,Yang}:
\begin{eqnarray}
|\Theta^{+}\rangle & = & \left| \overline{10};200,-1\frac{1}{2}
  \frac{1}{2} \right \rangle + d_{27}^{\Theta}\left
  |27;200,-1\frac{1}{2}\frac{1}{2}\right
\rangle+d_{\overline{35}}^{\Theta} \left|\overline{35};200,
  -1\frac{1}{2} \frac{1}{2} \right\rangle,
\cr
|n\rangle & = & \left| 8;1\frac{1}{2} - \frac{1}{2}, -1\frac{1}{2}
  \frac{1}{2} \right\rangle + c_{\overline{10}}^{N} \left|
  \overline{10}; 1\frac{1}{2}-\frac{1}{2}, -1\frac{1}{2} \frac{1}{2}
\right \rangle + c_{27}^{N}\left|27;1\frac{1}{2} -
  \frac{1}{2},-1\frac{1}{2} \frac{1}{2}\right\rangle.
\label{wvf}
\end{eqnarray}
The mixing coefficients are obtained from Eq.(\ref{wfc}) and yield as
follows:
\begin{equation}
c_{\overline{10}}^{N} = \sqrt{5}c_{\overline{10}},\;\;\;
c_{27}^{N} = \sqrt{6}c_{27},\;\;\;
d_{8}^{\Theta} = 0,\;\;\;
d_{27}^{\Theta} = \sqrt{\frac{3}{10}}d_{27},\;\;\;
d_{\overline{35}}^{\Theta} = \sqrt{\frac{1}{7}}d_{\overline{35}}
\end{equation}
with
\begin{eqnarray}
c_{\overline{10}}=-\frac{I_{2}}{15}\Big(\alpha+\frac{1}{2}\gamma\Big),
&  c_{27}=-\frac{I_{2}}{25}\Big(\alpha-\frac{1}{6}\gamma\Big), &
\cr
d_{8}=\frac{I_{2}}{15}\Big(\alpha+\frac{1}{2}\gamma\Big), &
d_{27}=-\frac{I_{2}}{8}\Big(\alpha-\frac{7}{6}\gamma\Big), &
d_{\overline{35}}=-\frac{I_{2}}{4}\Big(\alpha+\frac{1}{6}\gamma\Big).
\label{mixing_coeff}
\end{eqnarray}

Now, we are in a position to evaluate the baryonic matrix element
given in Eq.(\ref{general}) within the framework of the $\chi$QSM.
In general, the baryonic matrix element of a vector-current operator
$\mathcal{J}_\mu^\chi=i {\psi}^\dagger \gamma_\mu \lambda^\chi\psi$
can be expressed as the following correlation function in the
functional integral:
\begin{eqnarray}
\langle B^{\prime}(p^{\prime})|\mathcal{J}_\mu^\chi(0) |B(p)\rangle &
= &
\frac{1}{\mathcal{Z}} \lim_{T\to\infty} e^{-ip_{4}^{\prime}
  \frac{T}{2}+ip_{4}\frac{T}{2}} \int d^{3}x^{\prime}\,d^{3}x\;
e^{i{\bm p}\cdot{\bm x} - i{\bm p}^{\prime}\cdot{\bm x}^{\prime}}
\cr
 &  & \hspace{-3cm} \times\int\mathcal{D} \psi^{\dagger}\mathcal{D}\psi
\mathcal{D} U
J_{B^{\prime}}\left(\frac{T}{2},{\bm x}^{\prime}\right)\,\mathcal{J}_\mu^\chi
(0) J_{B}^{\dagger}\left(-\frac{T}{2},{\bm x}\right)\,
 e^{-\int d^{4}x\,\psi^\dagger D(U)\psi}.
\label{eq:corr}
\end{eqnarray}
with the baryonic current that consists of $N_c$ quarks:
\begin{equation}
J_B(x) = \frac{1}{N_c!}\epsilon_{i_1\cdots
  i_{N_c}}\Gamma_{JJ_3TT_3Y}^{\alpha_1\cdots\alpha_2}\psi_{\alpha_1i_1}
(x)\cdots \psi_{\alpha_{N_c}i_{N_c}}(x).
\end{equation}
Here, $\alpha_1\cdots\alpha_{N_c}$ denote spin-flavor indices,
whereas $i_1\cdots i_{N_c}$ represent color indices.

We can solve Eq.(\ref{eq:corr}) in the saddle-point approximation
justified in the large $N_c$ limit.  In this approximation and with
the help of the zero-mode quantization, the functional integral over
the chiral field turns out to be the integral over the rotational zero
modes.  Since we will consider the rotational $1/N_c$ corrections and
linear $m_{\mathrm{s}}$ corrections, we expand the quark propagators
in Eq.(\ref{eq:corr}) with respect to $\Omega$ and $\delta m$ to the
linear order and $\dot{T}_{z(t)}^{\dagger}T_{z(t)}$ to the zeroth
order.

Having carried out a tedious but straightforward
calculation~(see Refs.~\cite{Christov:1995vm,Kim_eleff} for details),
we finally can express the baryonic matrix element in
Eq.(\ref{general}) as a Fourier transform in terms of the
corresponding quark densities and collective wave-functions of the
baryons:
\begin{equation}
\langle B^{'}(p^{'})| \mathcal{J}_{\mu}^{\chi}(0) |B(p)\rangle =
\int d A \int d^{3}z\,\, e^{i{\bm q}\cdot{\bm z}}\,
\Psi_{B^{\prime}}^{*}(A) \mathcal{F}_{\mu}^\chi({\bm z})\Psi_{B}(A),
\label{eq:model}
\end{equation}
where $\Psi(A)$ denote the collective wave-functions and
$\mathcal{F}_{\mu}^{\chi}$ represents the quark densities
corresponding to the current operator $\mathcal{J}_\mu^\chi$.

Following the formalism presented above, we arrive at the final
expressions for the electromagnetic-like $nK^{+*} \to \Theta^+$ transition
form factors written as follows:
\begin{equation}
G_{E(M)}^{n\Theta}(Q^{2})  =
G_{E(M)}^{(\Omega^{0},m_{s}^{0})}(Q^{2}) +
G_{E(M)}^{(\Omega^{1},m_{s}^{0})}(Q^{2}) +
G_{E(M)}^{(\Omega^{0},m_{s}^{1}),\textrm{op}}(Q^{2}) +
G_{E(M)}^{(\Omega^{0},m_{s}^{1}),\textrm{wf}}(Q^{2}),
\label{eq:final}
\end{equation}
where the first term corresponds to the leading order
($\Omega^{0},m_{s}^{0}$), the second one to the rotational $1/N_{c}$
corrections ($\Omega^{1},m_{s}^{0}$) and the third and the last ones
to linear $m_{s}$ corrections coming from the operator and
wave-function corrections, respectively.

Since we have employed the large $N_c$ limit to solve the matrix
element of Eq.(\ref{eq:model}), we also should consider it
consistently in the relation of Eq.(\ref{vecqdef}).  Since the mass of
the $\Theta^+$ can be related to that of the neutron in the $\chi$QSM
by
\begin{equation}
M_{\Theta} = M_{n} + \frac{3}{2}\frac{1}{I_{2}} + \textrm{const}\cdot
m_{\mathrm{s}},
\label{mass}
\end{equation}
where $M_{n}$ and $I_{2}$ is proportional to $N_c$, i.e. $M_n,\,I_2
\sim\mathcal{O}(N_{c})$.  Thus, the second term is of order
$\mathcal{O}(N_{c}^{-1},m_{s}^{0})$ and the third
term of order $\mathcal{O}(N_{c}^{0},m_{s}^{1})$.  In the present work,
we take into account systematically only orders of
$\mathcal{O}(N_{c},m_{s}^{0})$, $\mathcal{O}(N_{c}^{-1},m_{s}^{0})$,
and $\mathcal{O}(N_{c}^{0},m_{s}^{1})$ for the quark densities
$\mathcal{F}_{\mu}^{\chi}({\bm z})$, while we consider
$\mathcal{O}(N^0_{c},m_{s}^{0})$ order for the translational modes.
Inserting Eq.(\ref{mass}) in Eq.(\ref{vecqdef}) and taking all terms
of order $\mathcal{O}(N^0_{c},m_{s}^{0})$, we end up with
\begin{equation}
{\bm q}^{2}\stackrel{N_{c}\to\infty}{=}Q^{2}.
\label{vecqnc}
\end{equation}
Thus, we can express the electromagnetic-like transition form factors as
functions of $Q^2$ as follows:
\begin{eqnarray}
G_{E}^{n\Theta}(Q^{2}) &=&\int\frac{d\Omega_{q}}{4\pi} \langle
\Theta(p^{\prime})| J_{V}^{4}(0)|n(p)\rangle = \int d^{3}z\,
j_{0}(|{\bm Q}||{\bm z}|) \mathcal{G}_{E}({\bm z}),
\label{model GE} \\
G_{M}^{n\Theta}(Q^{2}) &=& 3M_{n}\int\frac{d\Omega_{q}}{4\pi} \frac{q^{i}
  \epsilon^{ik3}}{ i\mid{\bm Q}\mid^{2}}\langle\Theta(p^{\prime})|
J_{V}^{k}(0)| n(p)\rangle\cr
&=& M_{n}\int d^{3}z\frac{j_{1}(|{\bm Q}||{\bm z}|)}{|{\bm Q}||{\bm z}|}
\mathcal{G}_{M}({\bm z}),
\label{model GM}
\end{eqnarray}
where $j_{0}$ and $j_{1}$ denote the usual spherical Bessel functions.
The $\mathcal{G}_{E}$ and $\mathcal{G}_M$ represent the electric-like
(magnetic-like) transition densities.  The final expressions for the
electromagnetic-like transition densities are expressed as follows:
\begin{eqnarray}
\mathcal{G}_{E}^{(m_{s}^{0})}({\bm z}) & = & -\frac{1}{2}\sqrt{
\frac{1}{15}} \mathcal{B}({\bm z}) + \frac{1}{2}\frac{3}{I_{1}}
\sqrt{\frac{1}{15}}\mathcal{I}_{1}({\bm z}),\cr
\mathcal{G}_{E}^{(m_{s}^{1}),\textrm{op}}({\bm z})
& = & -\frac{M_{8}}{I_{1}}\frac{1}{2}\sqrt{\frac{1}{5}}
\left[I_{1}\mathcal{K}_{1}({\bm z})-K_{1}\mathcal{I}_{1}({\bm z})
\right]+\left[M_{1}\sqrt{\frac{1}{15}}+\frac{M_{8}}{12}\sqrt{\frac{1}{5}}
\right]\mathcal{C}({\bm z}), \cr
\mathcal{G}_{E}^{(m_{s}^{1}),\textrm{wf}}({\bm z})
& = & \frac{1}{2}\left[c_{\overline{10}}\frac{1}{4}\sqrt{\frac{5}{3}}
-c_{27}\frac{3}{4}\sqrt{\frac{1}{15}}\right]\mathcal{B}({\bm z})
+\frac{1}{I_{1}}\frac{1}{2}\left[c_{\overline{10}}\frac{3}{4}\sqrt{
\frac{5}{3}}-c_{27}\frac{7}{4}\sqrt{\frac{3}{5}}\right]
\mathcal{I}_{1}({\bm z})\cr
 &  & +\frac{1}{I_{2}}\frac{1}{2}\left[c_{\overline{10}}\frac{9}{2}
\sqrt{\frac{5}{3}}+c_{27}\frac{5}{2}\sqrt{\frac{3}{5}}\right]
\mathcal{I}_{2}({\bm z}),\\
\mathcal{G}_{M}^{(m_{s}^{0})}({\bm z}) & = & -\frac{1}{2}\sqrt{
\frac{1}{15}}\left[\mathcal{Q}_{0}({\bm z})+\frac{1}{I_{1}}
\mathcal{Q}_{1}({\bm z})\right]+\frac{1}{I_{1}}\frac{1}{4}\sqrt{
\frac{1}{15}}\mathcal{X}_{1}({\bm z})+\frac{1}{I_{2}}\frac{1}{2}
\sqrt{\frac{1}{15}}\mathcal{X}_{2}({\bm z}),
\cr
\mathcal{G}_{M}^{(m_{s}^{1}),\textrm{op}}({\bm z}) & = & 3\left[M_{1}
\sqrt{\frac{1}{15}}+M_{8}\frac{1}{12}\sqrt{\frac{1}{5}}\right]
\mathcal{M}_{0}({\bm z})-M_{8}\frac{1}{12}\,\sqrt{\frac{1}{5}}
\left[3\mathcal{M}_{1}({\bm z})-\frac{K_{1}}{I_{1}}
\mathcal{X}_{1}({\bm z})\right]\cr
 &  & -M_{8}\frac{1}{6}\sqrt{\frac{1}{5}}\left[
3\mathcal{M}_{2}({\bm z})
-\frac{K_{2}}{I_{2}}\mathcal{X}_{2}({\bm z})\right],
\cr
\mathcal{G}_{M}^{(m_{s}^{1}),\textrm{wf}}({\bm z}) & = &
-\frac{1}{2}\left[c_{\overline{10}}\frac{1}{4}\sqrt{\frac{5}{3}}
-c_{27}\frac{7}{12}\sqrt{\frac{3}{5}}\right]\left[
\mathcal{Q}_{0}({\bm z})+\frac{1}{I_{1}}\mathcal{Q}_{1}({\bm z})
\right]\cr
 &  & -\frac{1}{I_{1}}\frac{1}{2}\left[c_{\overline{10}}\frac{1}{8}
\sqrt{\frac{5}{3}}-c_{27}\frac{1}{8}\sqrt{\frac{3}{5}}\right]
\mathcal{X}_{1}({\bm z})\cr
 &  & +\frac{1}{I_{2}}\frac{1}{2}\left[c_{\overline{10}}\frac{5}{8}
\sqrt{\frac{5}{3}}+c_{27}\frac{11}{24}\sqrt{\frac{3}{5}}\right]
\mathcal{X}_{2}({\bm z}).
\end{eqnarray}
 The explicit expressions for $\mathcal{B}({\bm z}),
\mathcal{I}_{i}({\bm z}),\mathcal{C}({\bm z}),\mathcal{K}_{i}({\bm z}),
\mathcal{Q}_{i}({\bm z}),\mathcal{X}_{i}({\bm z})$
and $\mathcal{M}_{i}({\bm z})$ can be found in Appendices
\ref{append:a} and \ref{append:b}.

Using the following relations~\footnote{The $\mathcal{B}$ represents
the baryon-number density multiplied by the number of valence
quarks, i.e. three for the neutron as well as for the $\Theta^+$
here.}:
\begin{equation}
\int d^{3}z\,\mathcal{B}({\bm z}) = 3,\;\; \frac{1}{I_{i}}\int
d^{3}z\, \mathcal{I}_{i}({\bm z}) = 1,\;\;
\frac{1}{K_{i}} \int d^{3}z\,\mathcal{K}_{i}({\bm z}) = 1,\;\;
\int d^{3}z\,\mathcal{C}({\bm z})=0,
\end{equation}
we can see that $G_E^{n\Theta}$ at $Q^2=0$ turns out to be just
proportional to $c_{\overline{10}}^{n}$, i.e. we have for
$G_E^{n\Theta}(0)$:
\begin{equation}
G_{E}^{n\Theta}(0) = \sqrt{15}c_{\overline{10}} =
\sqrt{3}c_{\overline{10}}^{n}.
\label{eq:admo1}
\end{equation}
This is a very interesting result, since it is a consequence of the
generalized Ademollo-Gatto theorem that we will prove in the next
Section.

In the large $N_c$ limit, the relations given in Eq.(\ref{eq:effGm})
can be simplified.  In this limit, we have
\begin{eqnarray}
E_{n}=\sqrt{M_{n}^{2}+\vec{p}_{n}^{2}} & \stackrel{
  N_{c}\to\infty}{=} &
E_{n}=M_{n}+\frac{\vec{p}_{n}^{2}}{2M_{n}}+\mathcal{O}(N_{c}^{-2}),\cr
\sqrt{\frac{E_{n}+M_{n}}{2M_{n}}} & \stackrel{ N_{c}\to\infty}{=}
& 1 + \mathcal{O}(N_{c}^{-2}),\cr
\frac{\vec{q}^{2}}{(M_{\Theta}+M_{n})(E_{n}+M_{n})} & \stackrel{
  N_{c}\to\infty}{=} &
\mathcal{O}(N_{c}^{-2})+\mathcal{O}(N_{c}^{-1},m_{s}^{1})+\cdots, \cr
\frac{q^{0}}{M_{\Theta}+M_{n}}=\frac{M_{\Theta}-E_{n}}{M_{\Theta}+M_{n}}
& \stackrel{ N_{c}\to\infty}{=} &
\mathcal{O}(N_{c}^{-2})+\mathcal{O}(N_{c}^{-1},m_{s}^{1})+\cdots.
\end{eqnarray}
Thus, the electromagnetic-like transition form factors derived in the
$\chi$QSM are simply related to the Dirac and Pauli transition form
factors as follows:
\begin{eqnarray}
G_{E}^{n\Theta}(Q^{2}) & = & F_{1}^{n\Theta}(Q^{2}),\label{eq:geqq} \\
G_{M}^{n\Theta}(Q^{2}) & = & F_{1}^{n\Theta}(Q^{2})+F_{2}^{n\Theta}(Q^{2}).
\label{eq:gmqq}
\end{eqnarray}
\section{A Generalization of the Ademollo-Gatto Theorem\label{sec:ag}}
In the present Section, we generalize the Ademollo-Gatto
theorem~\cite{Ademollo:1964sr,Cabibbo:2003cu} for the $K^{*}$ electric-like
transition form factor for the $ nK^{*+} \to \Theta^+$~\footnote{This
generalization was first done by M. V. Polyakov to whom the authors are
thankful.  In the Skyrme model a similar formalism has been described
in Ref.\cite{Park:1990}, though the generalized Ademollo-Gatto Theorem
was not discussed in Ref.\cite{Park:1990} in connection with the
transition from the $\Theta^+$ to the $nK^{*+}$.}.  The scalar current
$\overline{s}u$ in Eq.(\ref{divergence}) can be treated as an octet
representation, i.e. it behaves like $\mathrm{\overline{s}u} =
\overline{\Psi}(\lambda_4-i\lambda_5)\Psi/\sqrt{2} =(\kappa^-)^\dagger$, where
$\Psi$ denotes the quark field in the SU(3) fundamental
representation.  Here, we use the de Swart phase convention~\cite{deSwart}.
Thus, we obtain
\begin{equation}
\langle \Theta^+| \mathrm{\overline{s}u} | n \rangle \implies
\left[\langle \Theta^+ |\otimes \langle\kappa^-|\right] |n \rangle
= -\sqrt{\frac{2}{5}},
\label{eq:sc2}
\end{equation}
which is just one of the SU(3) Clebsch-Gordan
coefficients~\cite{deSwart}.  We can relate this matrix element
(\ref{eq:sc2}) to the wave-function
corrections $c_{\overline{10}}^n$: First let us consider the following
transition matrix element $\langle n |  \mathrm{\overline{s}s}
|n_{\overline{10}} \rangle$ in which the scalar current
$\mathrm{\overline{s}s} $ is a part of the SU(3)-symmetry breaking
Hamiltonian.  The mixing parameter between the baryon antidecuplet and
octet can be related to the octet part of the symmetry-breaking term
in the effective Hamiltonian: 
 \begin{equation}
c_{\overline{10}}^{n}=\frac{\langle n_{\overline{10}} |
H_{\textrm{sb}} | n \rangle}{M_{n} - M_{n_{\overline{10}}}}
=  \frac{-( m_s - \overline{m} )\langle
  n_{\overline{10}}  \mid \mathrm{\overline{s}s} \mid
  n\rangle}{M_{n}-M_{n_{\overline{10}}}}.
\end{equation}
The scalar current $ \mathrm{\overline{s}s}$ can be regarded as a
mixture of the singlet and octet currents,
i.e. $\mathrm{\overline{s}s} = \sqrt{\frac13}\eta_1
- \sqrt{\frac23} \eta_8$, so that we have 
\begin{equation}
  \label{eq:sc3}
\langle n_{\overline{10}} | \mathrm{\overline{s}s} | n \rangle \implies
\left[\left \langle n_{\overline{10}} \left|\otimes
      \left(\sqrt{\frac13} \langle\eta_1 |
- \sqrt{\frac23} \langle\eta_8|\right)
\right|n\right\rangle \right] =  \sqrt{\frac23}
\sqrt{\frac15}
\end{equation}
with $\eta_1=(\mathrm{\overline{u}u}+\mathrm{\overline{d}d} +
\mathrm{\overline{s}s})/\sqrt{3}$ and
$\eta_8=(\mathrm{\overline{u}u}+\mathrm{\overline{d}d} -2
\mathrm{\overline{s}s})/\sqrt{6}$.  Comparing Eq.(\ref{eq:sc3}) with
Eq.(\ref{eq:sc2}), we get the following relation:
\begin{equation}
\langle \Theta^+ \mid\mathrm{\overline{s}u} \mid n \rangle = -\sqrt{3}
\langle n_{\overline{10}} \mid\mathrm{\overline{s}s} \mid n \rangle.
\end{equation}
Thus, Eq.(\ref{divergence}) can be expressed for $Q^2=0$ as
\begin{equation}
 (M_{\Theta}-M_{n})F_{1}^{n\Theta}(0)  =
 \sqrt{3}c_{\overline{10}}^{n}(M_{n_{\overline{10}}}-M_{n}), 
\label{eq:admo2}
\end{equation}
which leads to the same expression as Eq.(\ref{eq:admo1}):
\begin{eqnarray}
F_1^{n\Theta}(0) &=& G_{E}^{n\Theta}(0)  = \sqrt{3}c_{\overline{10}}^{n}
\frac{M_{n_{\overline{10}}} - M_n}{M_{\Theta} - M_n}   \cr
&=& \sqrt{3}c_{\overline{10}}^{n}
\left(1+\mathcal{O}(m_{\mathrm{s}})\right).
\label{eq:admo3}
\end{eqnarray}
Since $c_{\overline{10}}^n$ is already known to be of linear order in
$m_{\mathrm{s}}$ from Eq.(\ref{mixing_coeff}), we conclude that
Eq.(\ref{eq:admo1}) is just a consequence of the generalized
Ademollo-Gatto theorem.  It asserts that the vector transition between
the baryon octet and antidecuplet receives the linear $m_{\mathrm{s}}$
corrections, while that within the baryon octet gets at most
$m_{\mathrm{s}}^2$ corrections, which is just called the original
Ademollo-Gatto theorem~\cite{Ademollo:1964sr}.  The present case is
similar to the case of the kaon semileptonic decay form factors: The
vector transition between the Goldstone bosons has also linear 
$m_{\mathrm{s}}$ corrections due to explicit chiral symmetry
breaking~\cite{Langacker:1973nf,Gasser:1984ux}.
\section{ Results and discussion for the transition form
  factors }
\begin{figure}[h]
\includegraphics[scale=0.5]{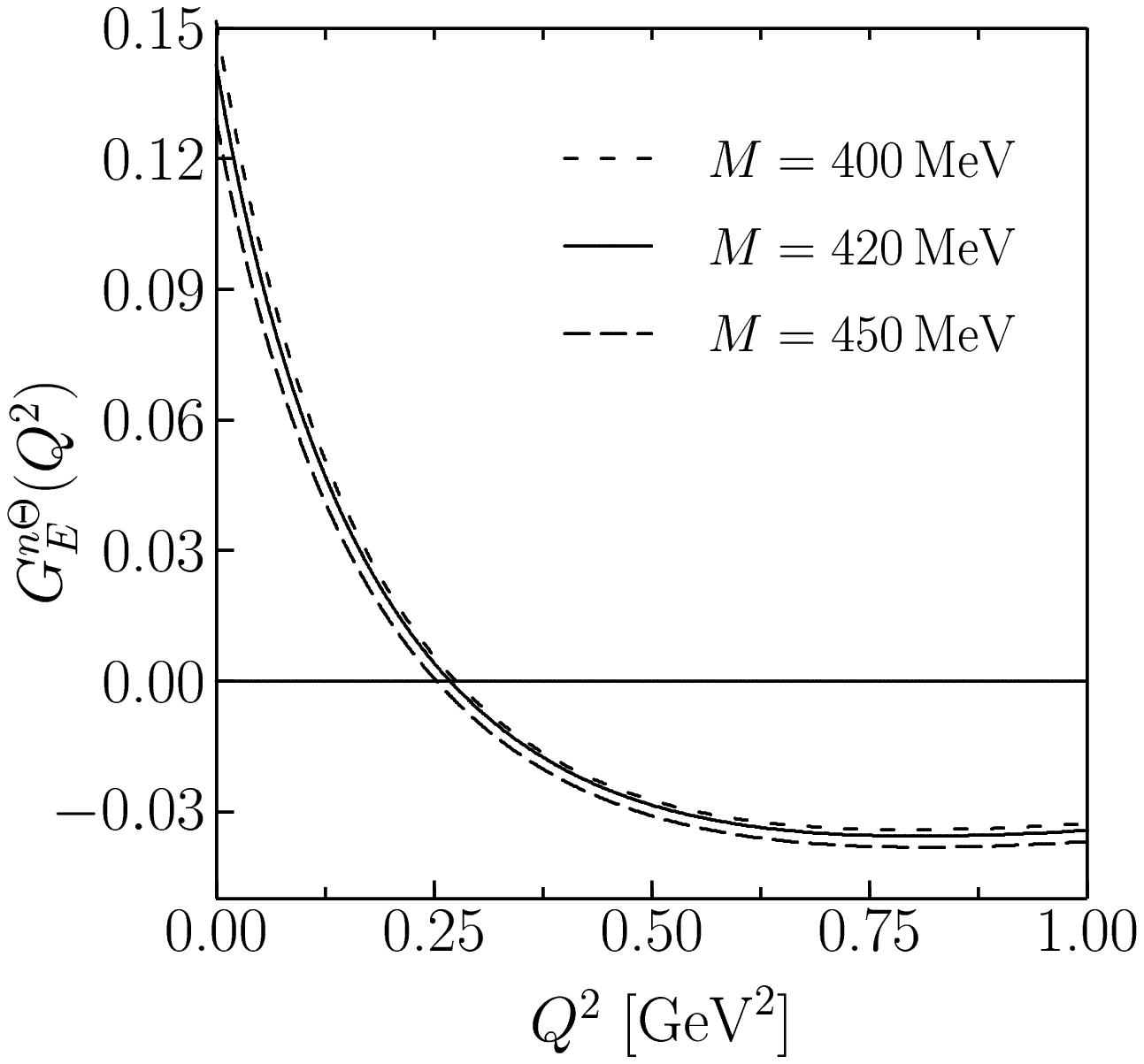}~~
\includegraphics[scale=0.5]{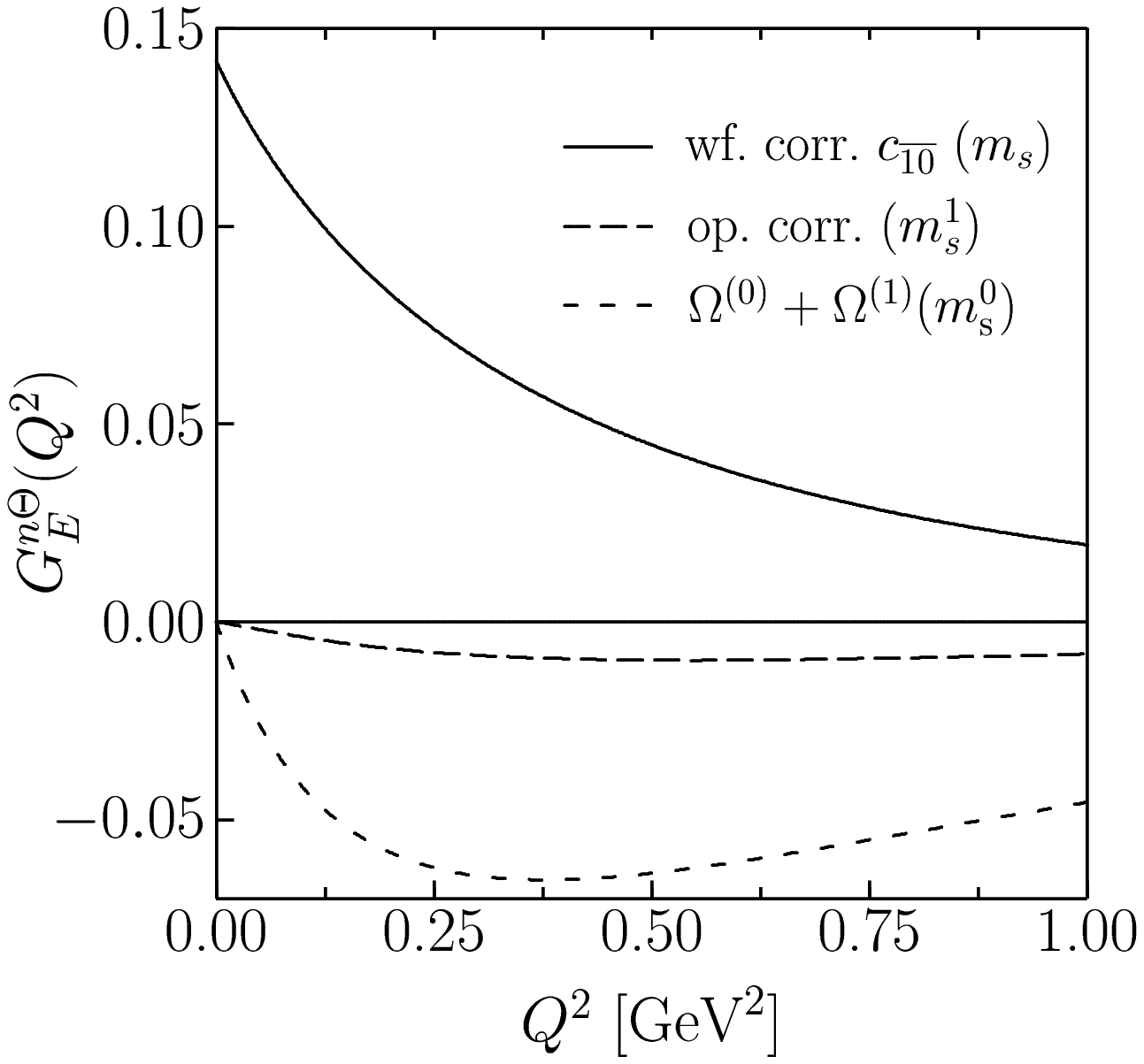}
\caption{Electric-Like transition form factor $G_E^{n\Theta}$ for the
  $nK^{*+} \to \Theta^{+}$ as a function of $Q^2$.  In the
left panel, the dependence of the form factor on the constituent quark
mass $M$ is drawn.  The solid curve depicts the form factor with
$M=420$ MeV, while the short-dashed and long dashed ones represent
that with $M=400$ and $M=450$ MeV, respectively.  In the right panel,
each contribution to the electric-like transition form factor is shown with
$M=420$ MeV.  The solid curve depicts the wave-function corrections,
while the long-dashed one draws the $m_{\mathrm{s}}$ corrections from
the operators.  The short-dashed one represents a sum of the
leadning-order and rotational $1/N_c$ corrections.}
\label{fig:1}
\end{figure}
We discuss now the results obtained from the present
work.  We refer to Refs.~\cite{Kim_eleff,Christov:1995vm} for a
detailed description of numerical methods.  Note that the only free
parameter of the $\chi$QSM is the constituent quark mass $M$.  In
general, most form factors of the baryons are insensitive to the
value of the $M$.  Usually, the $M$ was chosen to be $M=420$
MeV with which the best fit to many nucleon observables~\cite{
Christov:1995vm,SilvaKUG:2005,Kim_eleff,silva_data:2005}.  However,
as we will show in the next Section, the $K^*N\Theta^+$ coupling
constants are somewhat sensitive to the $M$, so that in the present
work we select the $M$ varying from 400 to 450 MeV.  The other
parameters of the model are the current nonstrange quark mass and the
cut-off parameter of the proper-time regularization: They are all
fixed for a given $M$ in such a way that mesonic properties, i.e. the
physical pion mass and decay constant, are exactly reproduced. The masses
of the current quarks are selected to be $\overline{m}=8$ MeV and
$m_{\rm s}=180$ MeV.  These parameters are known to
be the best to describe the mass-splitings between different baryon
representations.

\begin{figure}[h]
\includegraphics[scale=0.6]{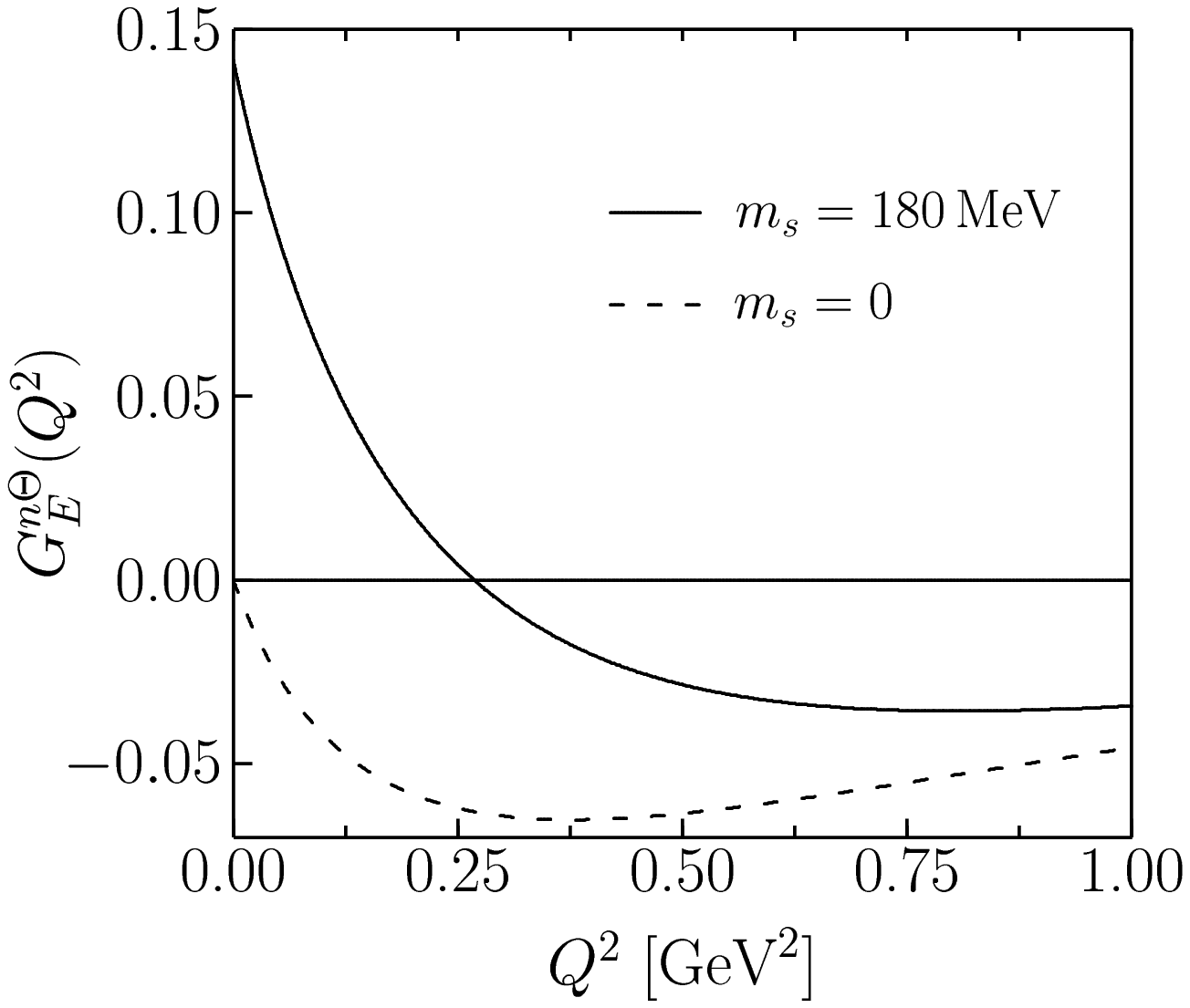}
\caption{Electric-Like transition form factor $G_E^{n\Theta}$ transition
for the $nK^{*+} \to \Theta^{+}$ as a function of $Q^2$.  The solid curve
depicte the form factor with the strange current quark mass
$m_{\mathrm{s}}=180$ MeV, while the dashed one draws that with the
$m_{\mathrm{s}}$ turned off. The constituent quark mass is taken to be
$420$ MeV.}
\label{fig:2}
\end{figure}

Figure~\ref{fig:1} draws the electric-like transition form factor
$G_{E}^{n\Theta}$ as a function of $Q^2$ for the $nK^*\to \Theta^+$.
As shown in the left panel of Fig.~\ref{fig:1}, its dependence on the
constituent quark mass $M$ is almost negligible.  In the right panel
of Fig.~\ref{fig:1}, we draw each contribution to the form factor.  It
is shown that the only contribution which survives at $Q^2=0$
is the wave-function correction with $c_{\overline{10}}$, which is a
consequence of the generalized Ademollo-Gatto theorem discussed in
Section~\ref{sec:ag}.  As $Q^2$
increases, however, all contributions become finite but are negative 
except for the wave-function corrections.  In particular, the
leading-order and rotational contributions decrease till $Q^2\simeq
0.4\,\mathrm{GeV}^2$, they start to increase rather mildly.  The
leading $m_{\mathrm{s}}$ correction decreases monotonically very
slowly.  In the lower $Q^2$ region, the wave-function corrections
are dominant, whereas the magnitude of the leading-order and
rotational $1/N_c$ contributions overcome those of the wave-function
corrections from $Q^2\simeq 0.4\,\mathrm{GeV}^2$.  Due to this fact,
the total electric-like transition form factor for the $\Theta^+\to n
K^*$ transition turns out to be negative around $Q^2\simeq
0.27\,\mathrm{GeV}^2$, as shown in Fig~\ref{fig:2}, where effects of
SU(3) symmetry breaking on the electric-like transition form factor
are drawn.

\begin{figure}[h]
\includegraphics[scale=0.6]{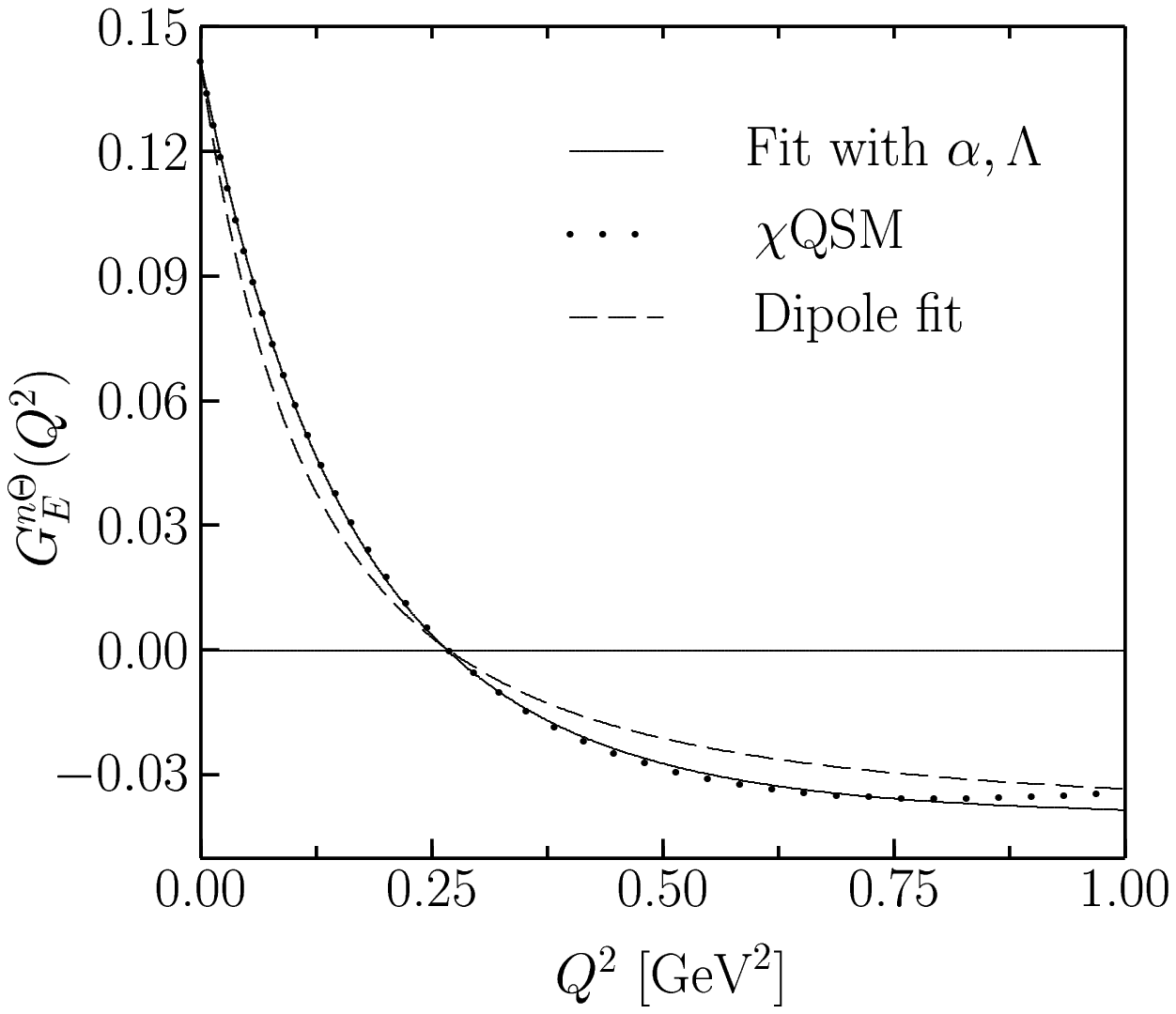}
\caption{The parameterization of the electric-like transition form factor
$G_E^{n\Theta}$ for the $nK^{*+} \to \Theta^{+}$ as a function of
$Q^2$.  The solid curve depicts the parameterized form factor
according to Eq.(\ref{eq:elfit}), whereas the dashed one is due to
the dipole-type fit of Eq.(\ref{eq:eldipol}).  The dotted one draws
the final result of the present work with $m_{\mathrm{s}}=180$ MeV and
$M=420$ MeV.}
\label{fig:3}
\end{figure}
As can be seen from Fig.~\ref{fig:2}, the electric-like transition form
factor $G_E^{n\Theta}$ behaves quite differently from the proton
electric form factor that can be well fitted by the dipole form
factor: $G_E^{p}(Q^2)=(1+Q^2/M_d^2)^{-2}$ with the dipole mass
$M_d=0.84\,\mathrm{GeV}$ in the lower $Q^2$
region~\cite{HydeWright:2004gh}.  Moreover, the proton electric form
factor is always positive through the whole region of $Q^2$.  However,
the eletric transition form factor $G_E^{n\Theta}$ turns negative from
around $Q^2\simeq 0.27\,\mathrm{GeV}^2$.  Thus, we want to
parameterize the present results for the $G_E^{n\Theta}$ in an
appropriate way.  It is of great use to make such a parameterization,
since it can be directly employed in relevant reaction calculations.
We introduce the following parameterization to reproduce the final
result shown in Fig.~\ref{fig:2}:
\begin{equation}
G_{E}^{n\Theta} (Q^{2}) =
\frac{G_{E}^{0} }{\left(1+\frac{Q^2}{\alpha
      \Lambda_E^2}\right)^{\alpha}}  + b,
\label{eq:elfit}
\end{equation}
where $\Lambda_E$ stands for the electric cut-off mass.  In order to
consider the fact that the $G_E^{n\Theta}$ becomes negative from
around $Q^2\simeq 0.27\,\mathrm{GeV}^2$, parameter $b$ is introduced
and is fitted to be $b=-0.04$.  Power $\alpha$ is rather sensitive to
the constituent quark mass, while  $\Lambda_E$ depends on it weakly.
In Table~\ref{tab:fit1}, we list the results for the fitted $b$,
$\Lambda_E$ and $\alpha$.
\begin{table}[h]
\caption{Fitted parameters and power of the parameterization given in
  \protect Eqs.(\ref{eq:elfit}) and (\ref{eq:eldipol}) as functions of the
  constituent quark mass $M$ in the range of $0\le Q^2 \le
  1\,\mathrm{GeV}^2$.}
\begin{tabular}{c|cccc}
\hline
$m_{s}=180$&
$ G^0_{E}$&
$\alpha$&
$\Lambda_E$&
$M_{d}$\tabularnewline
\hline
$M=400$&
$0.192$&
$6.33$&
$0.394$&
$0.487$\tabularnewline
$M=420$&
$0.182$&
$9.01$&
$0.402$&
$0.487$\tabularnewline
$M=450$&
$0.169$&
$37.6$&
$0.411$&
$0.480$\tabularnewline
\hline
\end{tabular}
\label{tab:fit1}
\end{table}
The parameteriztion of Eq.(\ref{eq:elfit}) reproduces the
present result very accurately, as shown in Fig.~\ref{fig:3}.
Instead of using the parameterization in Eq.(\ref{eq:elfit}), we could
also use the dipole form factor to fit the present result for the
$G_E^{n\Theta}$ as follows:
\begin{equation}
  \label{eq:eldipol}
G^{n\Theta}_{E}(Q^{2})
=\frac{ G^{0}_{E} }{\left(1+\frac{Q^2}{M_{d}^{2}}\right)^{2}}-0.04,
\end{equation}
where $M_d$ is called the dipole mass.  Though the shape of
the dipole form factor is qualitatively similar to the calculated
$G_E^{n\Theta}$ from the present model, it does not reproduce it
quantitatively.

\begin{figure}[h]
\includegraphics[scale=0.5]{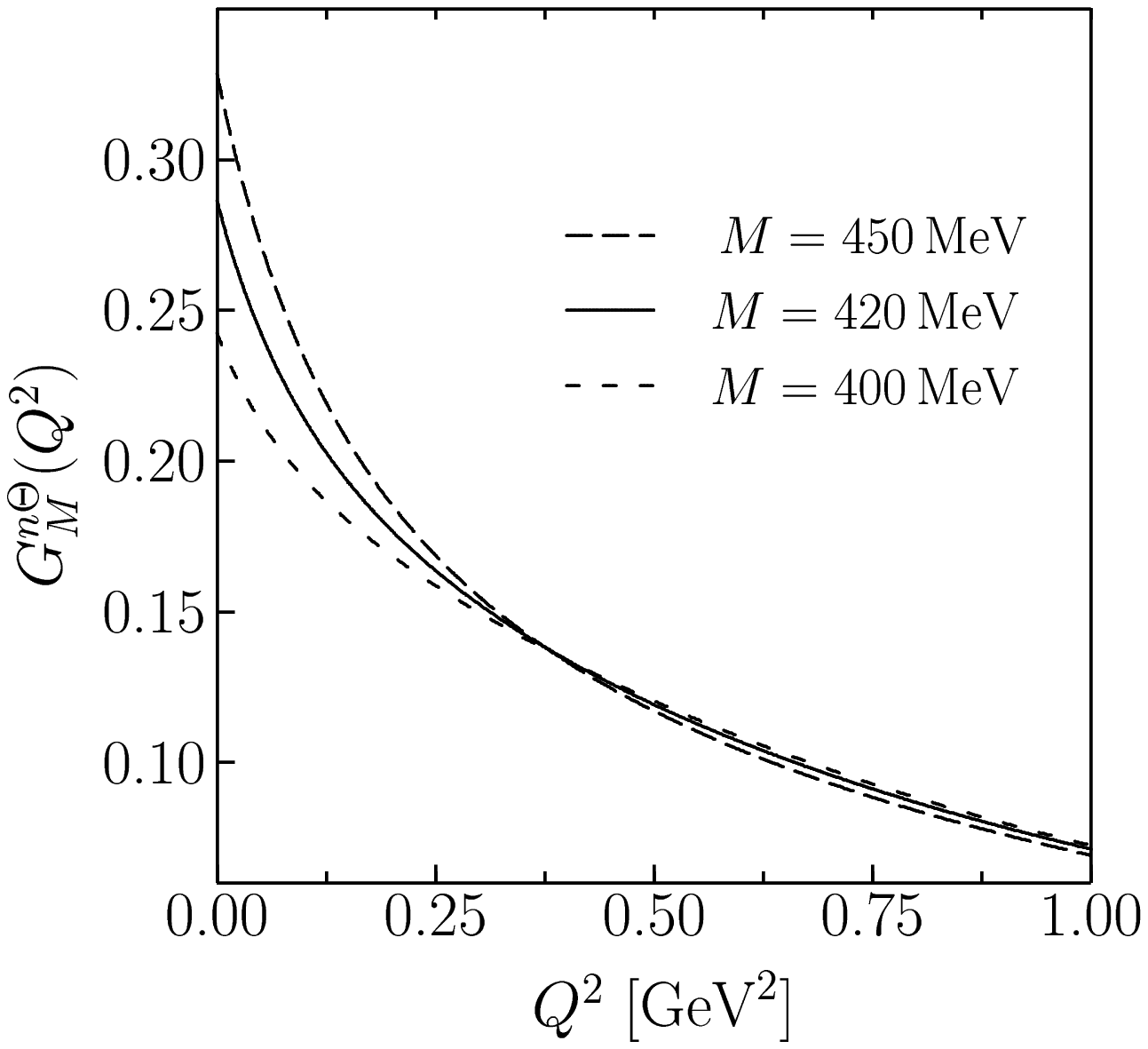}~~
\includegraphics[scale=0.5]{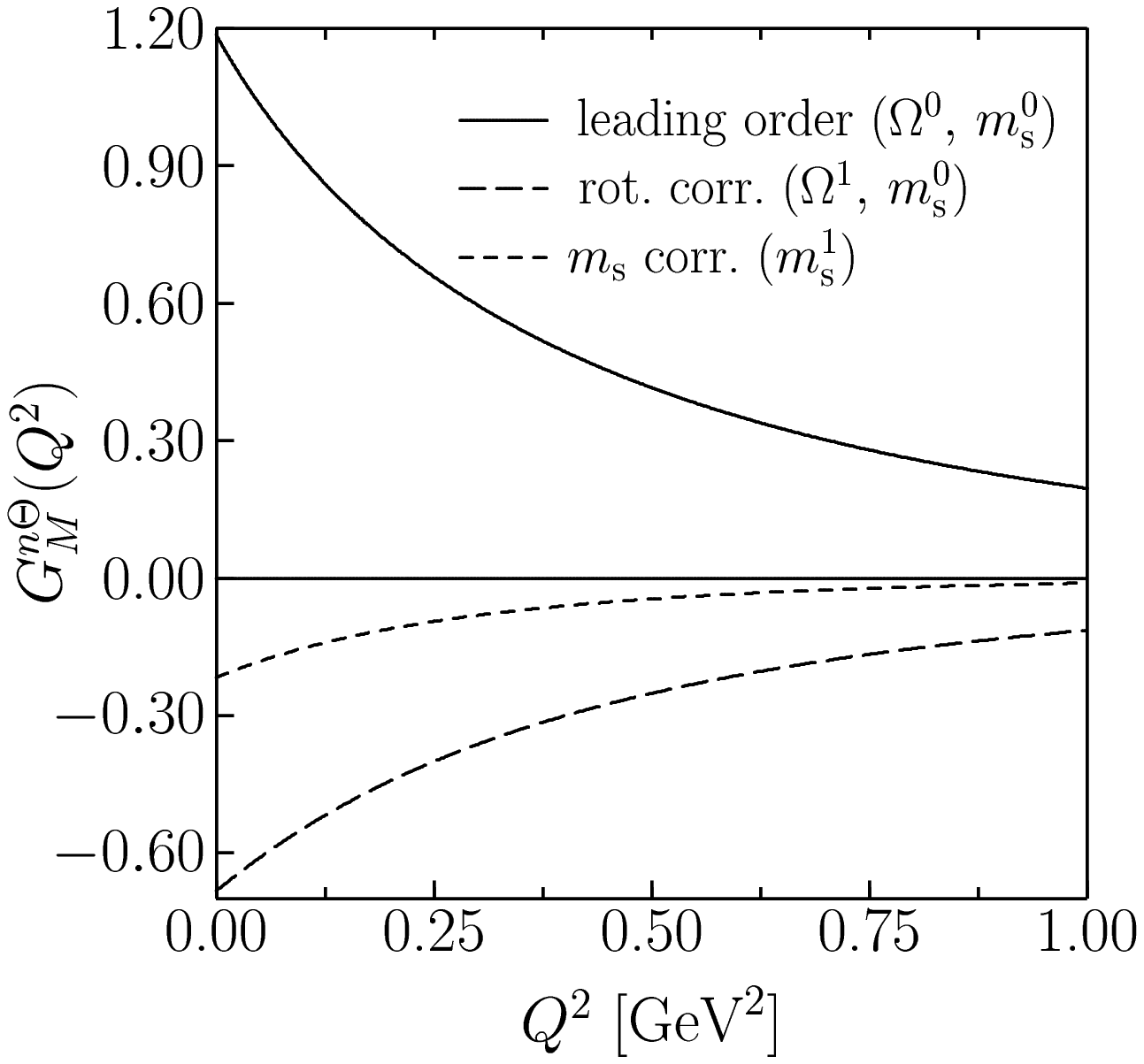}
\caption{Magnetic-Like transition form factor $G_M^{n\Theta}$ for the
  $nK^{*+} \to \Theta^{+}$ as a function of $Q^2$.  In the
left panel, the dependence of the form factor on the constituent quark
mass $M$ is drawn.  The solid curve depicts the form factor with
$M=420$ MeV, while the short-dashed and long dashed ones represent
that with $M=400$ and $M=450$ MeV, respectively.  In the right panel,
each contribution to the magnetic-like transition form factor is shown with
$M=420$ MeV.  The solid curve depicts the leading-order contributions,
while the long-dashed one draws the rotational $1/N_c$ corrections
from the operators.  The short-dashed one represents the
$m_{\mathrm{s}}$ corrections together with the wave-function ones.}
\label{fig:4}
\end{figure}
\begin{figure}[h]
\includegraphics[scale=0.6]{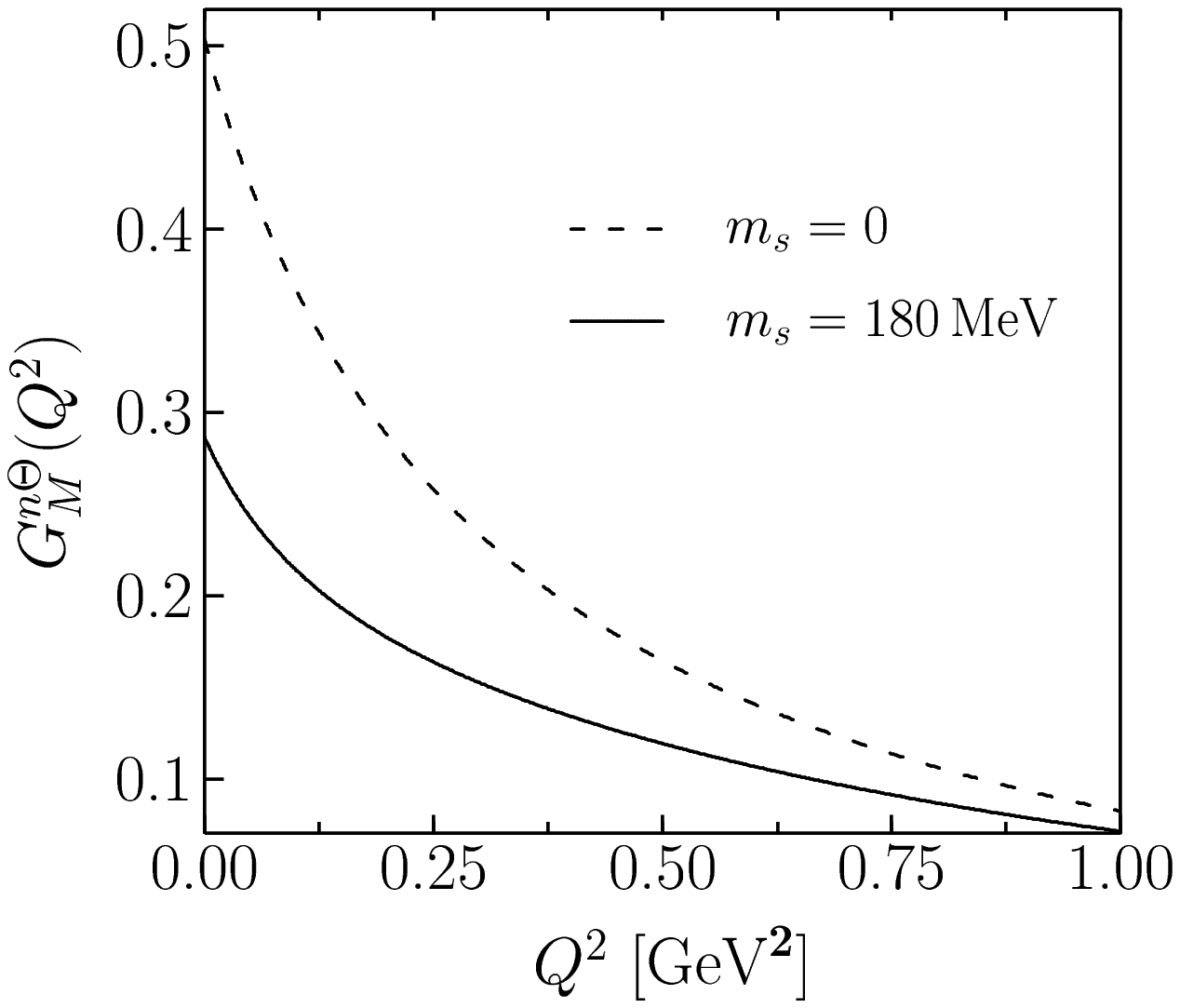}
\caption{Magnetic-Like transition form factor $G_M^{n\Theta}$.  Conventions
and parameters as in Fig.~\ref{fig:2}.}
\label{fig:5}
\end{figure}
We now discuss the results for the magnetic-like transition form factor
$G_M^{n\Theta}$.  In the left panel of Fig.~\ref{fig:4}, we draw the
$G_M^{n\Theta}$, varying the constituent quark mass from $M=400$ MeV
to $M=450$ MeV.  In contrast to the electric-like transition form factor,
the magnetic-like one depends noticeably on $M$ in the lower $Q^2$
region, as can be seen from the left panel of Fig.~\ref{fig:4}.
The magnetic-like transition form factor with $M=400$ MeV is approximately
$30\,\%$ smaller than that with $M=450$ MeV.  We will take this
difference as our model uncertainty.

We depict in the right panel of Fig.~\ref{fig:4} each contribution to
the magnetic-like transition form factor $G_M^{n\Theta}$.  Note that there
is a large cancellation between the leading-order contribution and the
rotational $1/N_c$ corrections.  This is again very different from the
case of the proton magnetic form factor to which the leading
order and rotational $1/N_c$ corrections contribute constructively.
As a result, though the linear $m_{\mathrm{s}}$ and wave-function
corrections look negligibly small, they turn out to be nonnegligible.
Thus, the effect of SU(3) symmetry breaking is of great significance
to describe the magnetic-like transition form factor for the $\Theta^+\to n
K^*$.  Because of this large cancelation, the magnetic-like transition
form factor becomes rather small, as shown in Fig.~\ref{fig:5}.  Since
the $m_s$ corrections contribute to the magnetic-like transition form
factor negatively, the $G_M^{n\Theta}$ in the SU(3) symmetric case
turns out to be almost $50\,\%$ larger than that with SU(3) symmetry
breaking.  Actually, this fact will play a very interesting role in
determining the $K^*$ coupling constants for the $\Theta^+$.

\begin{figure}[h]
\includegraphics[scale=0.6]{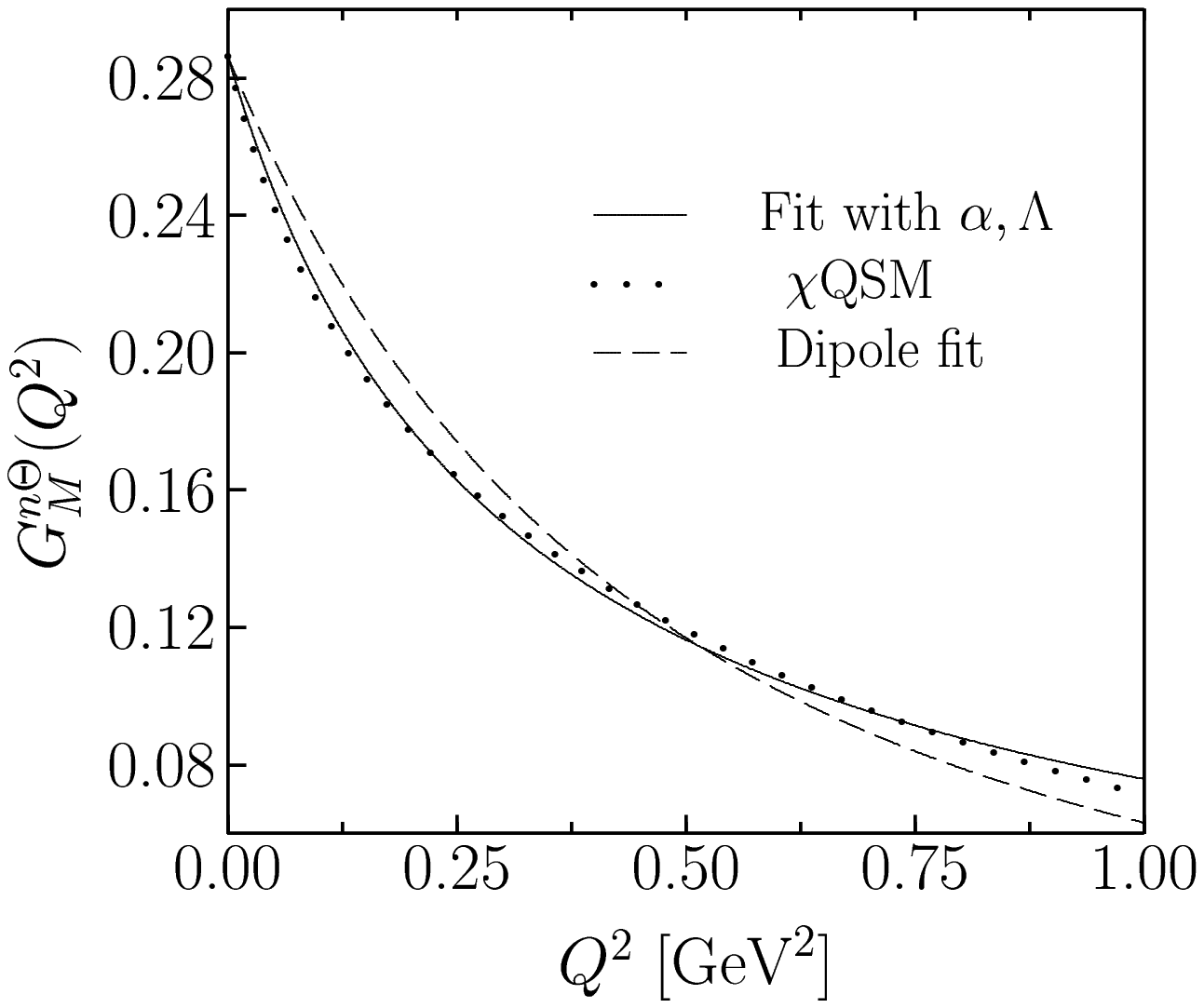}
\caption{The parameterization of the magnetic-like transition form factor
$G_M^{n\Theta}$ for the $nK^{*+} \to \Theta^{+}$ transition as a
function of $Q^2$.  Conventions as in Fig.~\ref{fig:3}.}
\label{fig:6}
\end{figure}

As already discussed in the case of the electric-like transition form
factor $G_E^{n\Theta}$, the parametrization of the magnetic-like transition
form factor is also of great interest for the same reason.
Empirically, the proton magnetic form factor is parameterized just in
the same way as the electric form factor: $G_M^{p}(Q^2)/\mu_p =
(1+Q^2/M_d^2)^{-2}$ with the common dipole mass
$M_d=0.84\,\mathrm{GeV}$ in the lower $Q^2$
region~\cite{HydeWright:2004gh}~\footnote{This common dipole form
factor deviates from the experimental data as $Q^2$
increases~\cite{Crawford:2006rz}.}.  The appropriate parameterization
for the magnetic-like transition form factor can be written as
\begin{equation}
  \label{eq:magfit}
G^{n\Theta}_{M}(Q^{2})=\frac{ G_{M}^{0}  }{\left(1
    +\frac{Q^2}{\alpha\Lambda_M^2}\right)^{\alpha}},
\end{equation}
where $\Lambda_M$ is the magnetic cut-off mass.  In Table~\ref{tab:fit2},
the results for fitting are listed for the case of $m_{\mathrm{s}}=0$
and $m_{\mathrm{s}}=180$ MeV, respectively.
\begin{table}[h]
\caption{Fitted parameters and power of the parameterization given in
  Eqs.(\ref{eq:magfit}) and (\ref{eq:magdipol}) as functions of the
  constituent quark mass $M$ in the range of $0\le Q^2 \le
  1\,\mathrm{GeV}^2$.}
\begin{tabular}{c|cccc}
\hline
$m_{s}=0$&
$ G^{0}_M  $&
$\alpha$&
$\Lambda_M$&
$M_{d}$\tabularnewline
\hline
$M=400$&
$0.485$&
$1.54$&
$0.558$&
$0.824$\tabularnewline
$M=420$&
$0.503$&
$1.49$&
$0.543$&
$0.808$\tabularnewline
$M=450$&
$0.516$&
$1.47$&
$0.531$&
$0.795$\tabularnewline
\hline
\end{tabular}~~
\begin{tabular}{c|cccc}
\hline
$m_{s}=180$&
$  G_{0} $&
$\alpha$&
$\Lambda$&
$M_{d}$\tabularnewline
\hline
$M=400$&
$0.242$&
$1.03$&
$0.688$&
$1.08$\tabularnewline
$M=420$&
$0.286$&
$0.851$&
$0.559$&
$0.942$\tabularnewline
$M=450$&
$0.328$&
$-$&
$-$&
$0.848$\tabularnewline
\hline
\end{tabular}
\label{tab:fit2}
\end{table}
In the case of the magnetic-like transition form factor, the parameters and
power are rather stable as $M$ varies.  The parameteriztion of
Eq.(\ref{eq:magfit}) reproduces very well again the result for the
$G_M^{n\Theta}$, as can be seen from Fig.~\ref{fig:6}.  If we use the
dipole form factor
\begin{equation}
  \label{eq:magdipol}
G_M^{n\Theta}(Q^{2})=
\frac{ G_{M}^{0}}{\left(1+\frac{Q^2}{M_d^2}\right)^{2}},
\end{equation}
the fitted result deviates from the calculated one, as shown in Fig.~\ref{fig:6}.

Since the effects of SU(3) symmetry breaking will be of great
significance in determining the $K^*n\Theta$ coupling constants,
we estimate the size of the $m_s$
corrections to the $G^{n \Theta}_E(0)$ and $G^{n \Theta}_M(0)$,
respectively, as follows:
\begin{eqnarray}
G^{n \Theta}_{E}: &  &
\frac{|G_{E}^{(m_{s}^{1})}|}{|G_{E}^{(\Omega^{0},m_{s}^{0})}| +
  |G_{E}^{(\Omega^{1},m_{s}^{0})}|+|G_{E}^{(m_{s}^{1})}|} =
\frac{0.14}{|0.39|+|-0.39|+|0.14|}=0.15
\cr
G^{n \Theta}_{M}: &  &
\frac{|G_{M}^{(m_{s}^{1})}|}{|G_{M}^{(\Omega^{0},m_{s}^{0})}| +
  |G_{M}^{(\Omega^{1},m_{s}^{0})}| + |G_{M}^{(m_{s}^{1})}|} =
\frac{|-0.16|}{|0.88| + |-0.51| + |-0.16|}=0.10,
\label{eq:mscor}
\end{eqnarray}
showing that the $m_s$ corrections turn out to be about $10-15\,\%$ to
the total form factors.

\begin{figure}[h]
\includegraphics[scale=0.5]{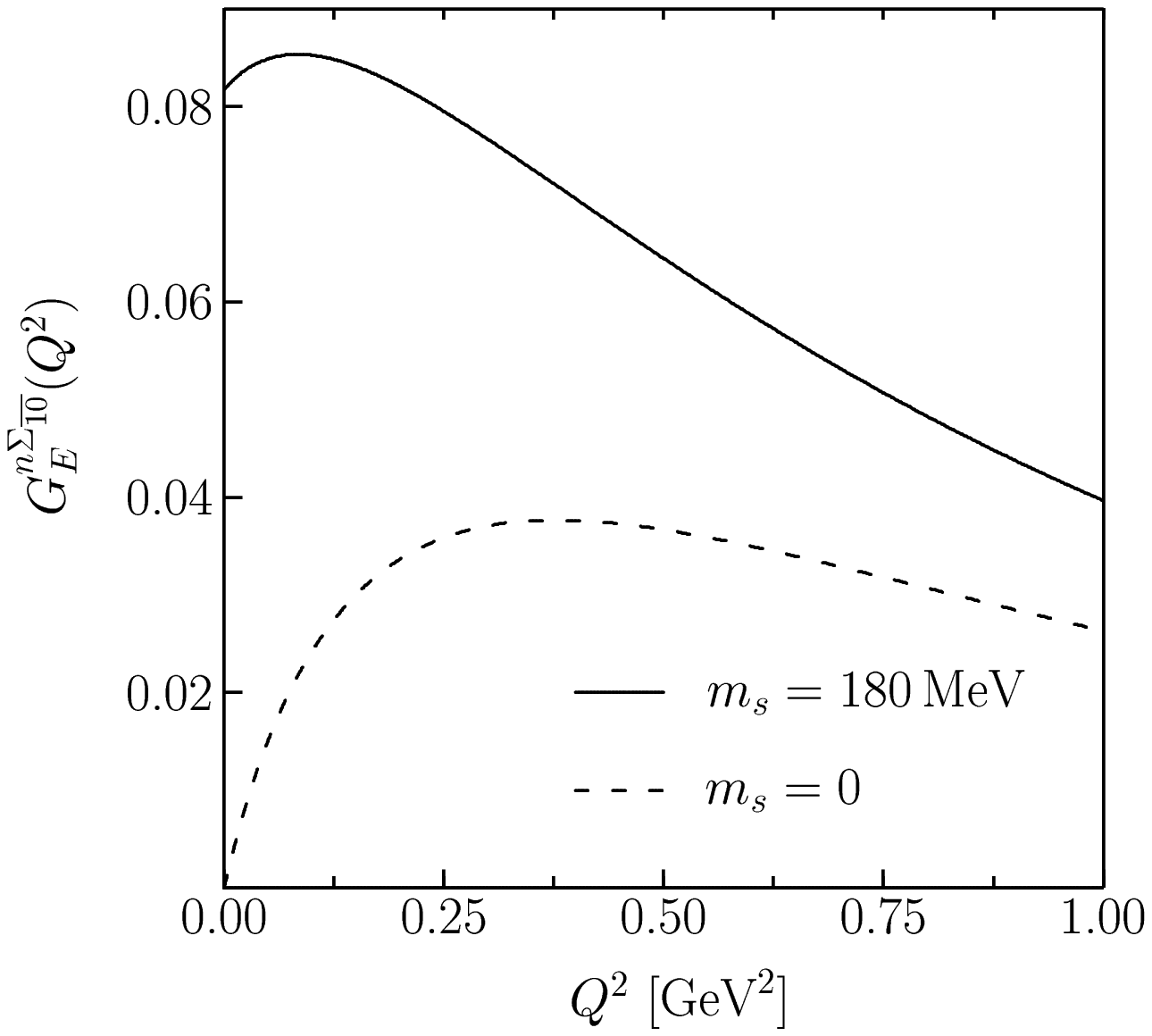}~~
\includegraphics[scale=0.5]{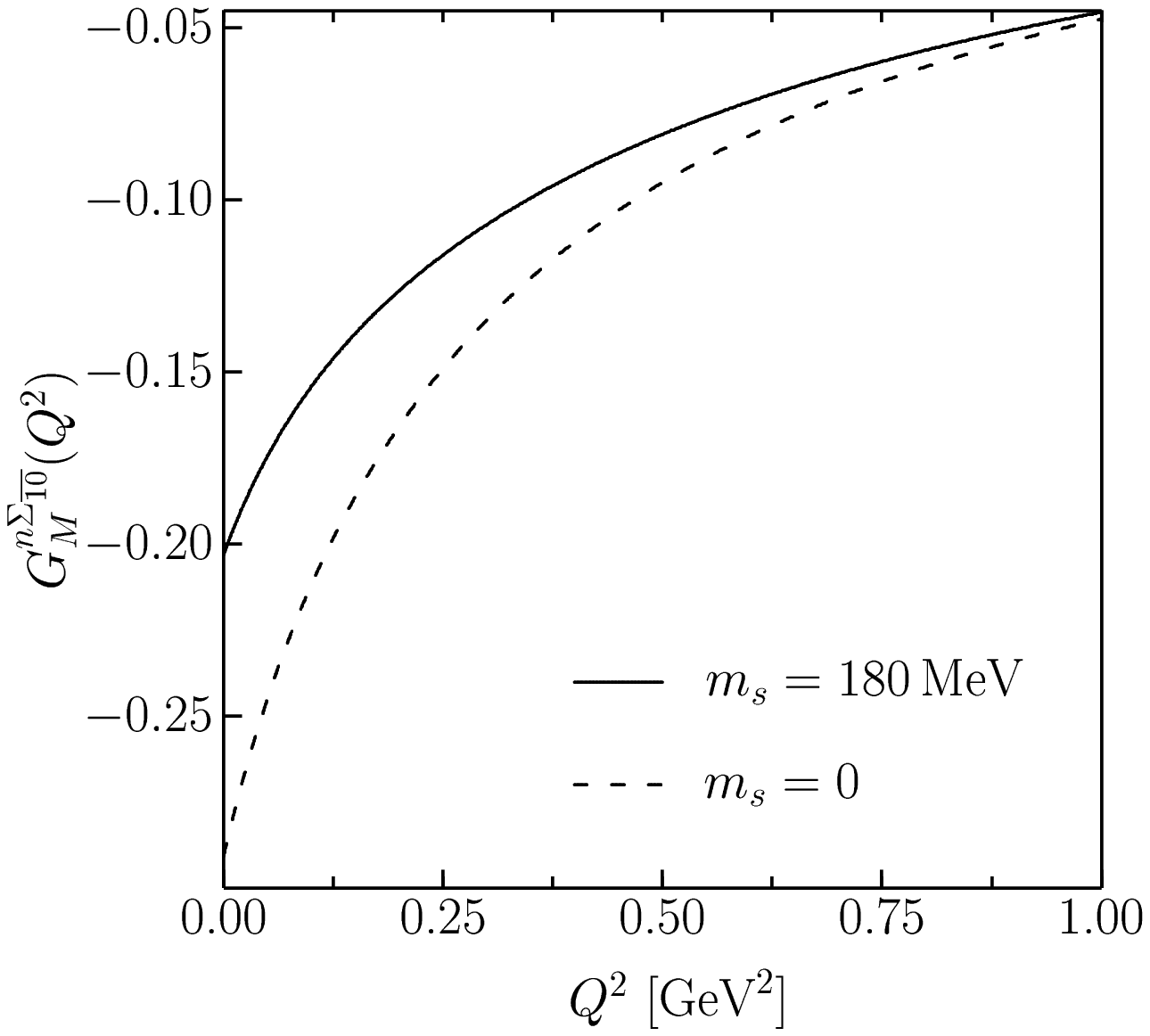}
\caption{Electric-Like and magnetic-like transition form factors for the $n\to
\Sigma^{-}_{\overline{10}}$ as functions of $Q^2$.  The solid curves
depict those with $m_{\mathrm{s}}=180$ MeV, whereas the dashed ones
draw those with $m_{\mathrm{s}}=0$.  The constituent quark mass is
taken to be $M=420$ MeV.  At $Q^2=0$, we have the following values:
$G_E^{n\Sigma_{\overline{10}}}(0)=0$ and
$G_M^{n\Sigma_{\overline{10}}}(0)=-0.291$ without $m_{\mathrm{s}}$
corrections: $G_E^{n\Sigma_{\overline{10}}}(0)=0.082$ and
$G_M^{n\Sigma_{\overline{10}}}(0)=-0.203$ with $m_{\mathrm{s}}$.} 
\label{fig:7}
\end{figure}

For completeness, we also present the results for the electric-like and
magnetic-like $nK^{*-} \to \Sigma^{-}_{\overline{10}}$ transition form
factors in Fig.~\ref{fig:7}.

\section{Results and Discussion for the $K^*$ coupling constants}
In order to estimate the $K^{*}n\Theta$ coupling constants from the
electric-like and magnetic-like transition form factors, we use the VMD that has
been discussed in Section~\ref{sec:vmd}.  In the large $N_c$ limit, we
have shown that the electromagnetic-like transition form factors can be
simply related to the Dirac and Pauli transition form factors in
Eqs.(\ref{eq:geqq}) and (\ref{eq:gmqq}).  Putting them into
Eq.(\ref{eq:gf}) at $Q^2=0$, we obtain the following relations for the
$K^*n\Theta$ vector coupling constant $g_{K^*n\Theta}$ and tensor one
$f_{K^*n\Theta}$, respectively:
\begin{equation}
g_{K^{*}n\Theta}=f_{K^{*}}\, G_E^{n \Theta}(0),\;\;
f_{K^{*}n\Theta}=f_{K^{*}}\,\left[G_{M}^{n \Theta}(0)-G_{E}^{n
    \Theta}(0)\right]
\label{eq:gKNT fKNT}
\end{equation}
In Table~\ref{tab:em3}, we summarize the values of the electromagnetic-like
transition form factors at $Q^2=0$.  The results for the
$K^{*+}n\Theta^+$ coupling constants are listed in
Table~\ref{tab:final4}.
\begin{table}[h]
\caption{The values of the electromagnetic-like transition form factors for
the $nK^{*+}\to\Theta^+$ at $Q^2=0$ with and without $m_{\mathrm{s}}$
corrections.  The constituent quark mass $M$ varies from $400$ MeV to
$450$ MeV.  The final results are those for $M=420$ MeV and
$m_{\mathrm{s}}$ MeV.}
\begin{tabular}{c|cc}
\hline
$m_{s}=0$&
$G^{n \Theta}_{E}(0)$&
$G^{n \Theta}_{M}(0)$\tabularnewline
\hline
$M=400$ &
$0$&
$0.486$\tabularnewline
$M=420$&
$0$&
$0.504$\tabularnewline
$M=450$&
$0$&
$0.516$\tabularnewline
\hline
\end{tabular}~~~~~
\begin{tabular}{c|cc}
\hline
$m_{s}=180$&
$G^{n \Theta}_{E}(0)$&
$G^{n \Theta}_{M}(0)$\tabularnewline
\hline
$M=400$&
$0.152$&
$0.243$\tabularnewline
$M=420$&
$0.142$&
$0.286$\tabularnewline
$M=450$&
$0.129$&
$0.328$\tabularnewline
\hline
\end{tabular}
\label{tab:em3}
\end{table}
\begin{table}
\caption{The results for the $K^*n\Theta^+$ coupling constants at
$Q^2=0$ with and without $m_{\mathrm{s}}$ corrections.  The
constituent quark mass $M$ varies from $400$ MeV to
$450$ MeV.  The final results are those for $M=420$ MeV and
$m_{\mathrm{s}}$ MeV.}
\begin{tabular}{c|cc}
\hline
$m_{s}=0$&
$g_{K^{*}n\Theta}$&
$f_{K^{*}n\Theta}$\tabularnewline
\hline
$M=400$&
$0$&
$2.80$\tabularnewline
$M=420$&
$0$&
$2.91$\tabularnewline
$M=450$&
$0$&
$2.97$\tabularnewline
\hline
\end{tabular}~~~~~
\begin{tabular}{c|cc}
\hline
$m_{s}=180$&
$g_{K^{*}n\Theta}$&
$f_{K^{*}n\Theta}$\tabularnewline
\hline
$M=400$&
$0.87$&
$0.53$\tabularnewline
$M=420$&
$0.81$&
$0.84$\tabularnewline
$M=450$&
$0.74$&
$1.16$\tabularnewline
\hline
\end{tabular}
\label{tab:final4}
\end{table}
Since the $K^*$ vector coupling constant $g_{K^*n\Theta}$ depends only
on the electric-like transition form factor $G_E^{n \Theta}$, it is rather
stable as $M$ varies, as shown in the left panel of Fig.~\ref{fig:1}.
However, the situation is slightly more complicated for the tensor
coupling constant $f_{K^{*}n\Theta}$.  Since the $G_E^{n\Theta}$ vanishes
in the SU(3) symmetric case, the tensor coupling constant is solely
determined by the magnetic-like transition form factor $G_M^{n\Theta}$, so
that it is still stable for the $M$ due to the fact that the
$G_M^{n\Theta}$ is insensitive to the $M$ without $m_{\mathrm{s}}$
corrections, as shown in the left panel of Table~\ref{tab:final4}.  When
the $m_{\mathrm{s}}$ corrections are switched on, the $f_{K^*n\Theta}$
gets a negative contribution from the $G_E^{n\Theta}$.  Thus, the
size of the tensor coupling constant $f_{K^*n\Theta}$ drastically
decreases and moreover it becomes very sensitive to the constituent
quark mass, as can be seen from the right panel of
Table~\ref{tab:final4}.

Because of the facts that the $G_E^{n\Theta}$ is finite  only with SU(3)
symmetry breaking due to the Ademollo-Gatto theorem and that the
$G_M^{n\Theta}$ is lessened by about $50\,\%$ due to the
$m_{\mathrm{s}}$ corrections, the effects of SU(3) symmetry breaking
play an essential role in determining the $K^*n\Theta$ coupling
constants, though the $m_{\mathrm{s}}$ corrections contribute to the
electromagnetic-like transition form factors by about $10-15\,\%$ as shown
in Eq.(\ref{eq:mscor}).

Considering the model uncertainty due to the dependence of the
electromagnetic-like form factors on the constituent quark mass, we finally
present the values of the vector and tensor $K^*n\Theta^+$ coupling
constants as follows:
\begin{equation}
  \label{eq:final3}
g_{K^{*}N\Theta}=0.74-0.87,\;\;
f_{K^{*}N\Theta}=0.53-1.16.
\end{equation}
Similarly, we also can compute the 
$K^{*-}n\to\Sigma^-_{\overline{10}}$ 
coupling constants as follows:
\begin{equation}
g_{K^{*} n \Sigma_{\overline{10}} } = 0.42-0.50,\;\;\;
f_{K^{*} n  \Sigma_{\overline{10}} }= (-1.59) - (-1.67).
\end{equation}
The smaller uncertainty in the case of
$K^{*-}n\to\Sigma^-_{\overline{10}}$ compared to that in
$K^*n\to\Theta^+$ is a consequence of the fact that in the former case
the form factors have different signs whereas in the latter case both
form factors are positive, yielding a cancelation effect due to
Eq.(\ref{eq:gKNT fKNT}).  At this point we have to mention that a
mixing of the $\Sigma^-_{\overline{10}}$ with radial excitations of
the $\Sigma^-_{8}$ has been neglected since in the present formalism 
radial excitations are not computable.

It is also very interesting to see how strong the $K^*n\Theta$
coupling constants are in comparison with the octet transition
coupling constants, for example, with the octet $\Lambda \to pK^*$
transition.  Within the same framework as the present work, we obtain
the electromagnetic-like transition form factors for the $\Lambda \to pK^*$
as follows: $G_E^{\Lambda p}(0)= 1.22$ and $G_M^{\Lambda p}(0)= 3.00$
in the flavor $SU(3)$ symmetric case.  Note that in the case of the
baryon octet there is no such large cancelation between the leading
order and the rotational $1/N_c$ corrections.  Moreover, the $m_s$
corrections are rather small. Thus, using Eq.(\ref{eq:gKNT fKNT}) for
the $\Lambda \to pK^*$ process, we get the  $K^*p\Lambda$ coupling
constants as follows:
\begin{equation}
  \label{eq:3}
 |g_{K^{*}\Lambda p}|=6.97,\;\;\;|f_{K^{*}\Lambda p}|=10.15
\end{equation}
which are comparable to those used in the Nijmegen baryon-baryon
interaction~\cite{Stoks:1999bz}.  Thus, we conclude that the
$K^*n\Theta$ coupling constants obtained in
the present work is very tiny, compared to the baryon octet ones.
Moreover, this small values of the $K^*n\Theta$ coupling
constant is in agreement with the conclusion of the recent
measurement~\cite{Miwa:2007xk}.

In Refs.~\cite{Oh:2003kw,Nam:2004xt}, the tensor coupling
constant $f_{K^*N\Theta}$ is set equal to zero, while the vector one
$g_{K^*N\Theta}$ is taken to be proportional to the $g_{KN\Theta}$,
since there is no theoretical guideline for it.
Assuming that the $\Theta^+$ has positive parity,
Ref.~\cite{Oh:2003kw} takes $g_{K^*N\Theta}=\sqrt{3}g_{KN\Theta}\simeq
1.7$ and Ref.~\cite{Nam:2004xt} employs $g_{K^*N\Theta}$ being varied
between $-g_{KN\Theta}/2$ and $g_{KN\Theta}/2$, i.e. $-1.9\le
g_{K^*N\Theta} \le 1.9$ ($\Gamma_\Theta\simeq 15$ MeV).
The values of the vector coupling constant $g_{K^*N\Theta}$ used in
Refs.~\cite{Oh:2003kw,Nam:2004xt} are at least about two times larger
than the present result.  It implies that if the present results were
employed in the calculation of the $\Theta^+$ photoproduction one
would have much smaller results for its production cross sections.

In Ref.~\cite{Kwee:2005dz}, the very small decay width of the
$\Theta^+$ and flavor SU(3) symmetry being assumed, the nonzero tensor
coupling constant $f_{K^*N\Theta}=1.1$ is obtained from the value of
the pseudoscalar coupling constant $g_{KN\Theta}\approx 1.056$, while
the vector coupling constant is dropped out.  The value of the
$f_{K^*N\Theta}$ in Ref.~\cite{Kwee:2005dz} is comparable to our value
with $M=450$ MeV, which is the largest one in the present work.
Ref.~\cite{Kwee:2005dz} predicted the total cross sections for the
$K^+$ photoproduction: $\sigma_{tot}(\gamma n\to
K^{-}\Theta^{+})<1\,\textrm{nb}$ and $\sigma_{tot}(\gamma
p\to\bar{K}^{0}\Theta)<0.22\,\textrm{nb}$ which are consistent with
the recent measurements by the CLAS
collaboration~\cite{DeVita:2006ha,McKinnon:2006zv}.  Note that, however,
Ref.~\cite{Kwee:2005dz} did not consider the vector coupling constant
and assumed SU(3) symmetry.

Ref.~\cite{Azimov:2006he} extracted the $f_{K^{*}\Theta N}$, assuming
again flavor $SU(3)$ symmetry and using their value of the $n\to
n_{\overline{10}}$ transition magnetic moment~\cite{Azimov:2005jj}.
Ref.~\cite{Azimov:2006he} obtained the tensor coupling constants as 
follows: $f_{K^{*}\Theta N}=|1.10-3.14|$ and $f_{K^{*-} p \Sigma^{*0}
}=|0.48-1.28|$. In the present work, we obtain for the tensor coupling 
constant for  $K^{*-}p\to\Sigma^0_{\overline{10}}$ the value$f_{K^{*-}
p  \Sigma^{*0} }=-1.17$ in the $SU(3)$ symmetrics case. 
Since SU(3) symmetry was assumed, 
Ref.~\cite{Azimov:2006he} was not able to extract the vector coupling
constant.  In the $\chi$QSM, we are also able to calculate the
$n\to n_{\overline{10}}$ transition magnetic moment within the same
framework with the same set of parameters. In the SU(3) symmetric
case, we obtain $\mu_{n_{\overline{10}}n}=0.30$ n.m. for the
transition magnetic moment.  Following the formalism of
Ref.~\cite{Azimov:2006he}, we get $f_{K^{*}\Theta N}=2.56$,
which is compatible to the value $f_{K^{*}\Theta N}=2.91$ presented in
Table~\ref{tab:final4}.  The range of the $f_{K^{*}\Theta N}$ given in
Ref.~\cite{Azimov:2006he} is larger than the present result of
Eq.(\ref{eq:final3}).  This may be due to the fact that the effects of
SU(3) symmetry breaking have been neglected in
Ref.~\cite{Azimov:2006he}.

If one uses the present results in reaction calculations like in
Refs.\cite{Oh:2003kw,Nam:2004xt,Kwee:2005dz}, one will probably
further bring down the total cross sections for the
$\Theta^{+}$-photoproduction.  The production cross sections will turn
out to be even smaller than those estimated in \cite{Kwee:2005dz}.
Thus, we want to emphasize that the $K^*N\Theta$ coupling constants
obtained in this work are consistent with the recent findings of CLAS
and KEK~\cite{DeVita:2006ha,McKinnon:2006zv,Miwa:2007xk}. 
\section{Summary}
In the present work, we have investigated the electromagnetic-like transition
form factors for the $n\to\Theta^{+}$ transition within the framework
of the chiral quark-soliton model. We have assumed isospin
symmetry and have incorporated the symmetry-conserving quantization.
The rotational $1/N_{c}$ and strange quark mass $m_{s}$ corrections
were taken into account.  The value of the constituent quark mass,
the only free parameter in the chiral quark-soliton model, was chosen
in the range of $M=400-450$MeV for which many properties of the octet
and decuplet baryons are for many years known to be well
reproduced. In accordance with previous calculations, we have selected
$M=420$MeV as the central value and take the dependence of the results
on the constituent quark mass as our model uncertainty.  We first have
studied the electromagnetic-like transition form factors for the
$nK^{*+}\to\Theta^{+}$ process.

We have found that the electric-like transition form factors for this
process has a nonvanishing value at $Q^{2}=0$ solely due to the SU(3)
wave-function corrections, which is a consequence of the generalized
Ademollo-Gatto theorem derived in this work.  Since the wave-function
corrections are dominant in the lower $Q^2$ region and all other
contributions turn out negative, the electric-like transition form factor
behaves quite differently, compared to usual form factors such as the
proton electric form factors.  As a result,
the electric-like transition form factors for the $nK^{*}\to\Theta^{+}$
changes sign at $Q^{2} \approx 0.3\, \textrm{GeV}^2$. Since the
electric-like transition form factor at $Q^2=0$ is determined only by the
wave-function corrections, we have obtained a very small value of the
vector meson $K^{*}$ coupling constant for the $n\to\Theta^{+}$
transition. In the chiral quark-soliton model we have the range of
this coupling constant as: $g_{K^{*}N\Theta}=0.74-0.87$.  The present
result is much smaller than those used in reaction
calculations~\cite{Oh:2003kw,Nam:2004xt}.

When it comes to the magnetic-like transition form factor for the
$nK^{*}\to\Theta^{+}$, we have found that there is a large
cancelation between the leading order and rotational $1/N_{c}$
corrections, which leads to the tiny magnetic-like transition form
factor. Thus, even though the $m_{s}$ corrections
are not large, they are still very essential due to that large
cancelation. Because of this fact, the tensor coupling constant for
the $nK^{*}\to\Theta^{+}$ turns out to be rather small, i.e. it lies
in the range of $f_{K^{*}N\Theta}=0.53-1.16$.  It is even smaller than
those of Refs.\cite{Kwee:2005dz,Azimov:2006he}.  It indicates that
with the $K^*n\Theta$ coupling constants from the present results,
the total cross sections for the $\gamma n\to K^{*}\Theta$ will appear
much smaller than those obtained in the reaction
calculations~\cite{Oh:2003kw,Nam:2004xt,Kwee:2005dz}.
These very small $K^*n\Theta$ coupling constants are consistent with
the recent CLAS and KEK
measurements~\cite{DeVita:2006ha,McKinnon:2006zv,Miwa:2007xk}.

In conclusion, we have learned two significant features for the
$nK^{*+}\to\Theta$ coupling constants from the present study: Firstly,
the $nK^{*+}\to\Theta$ vector coupling constant does not vanish at the
point $Q^{2}=0$ when the effects of SU(3) symmetry breaking are taken
into account.  Secondly, the tensor coupling constant gets even
smaller with the $m_{s}$ corrections taken into account. Altogether
those coupling constants are quite sensitive to the effects of SU(3)
symmetry breaking and seem to be reduced considerably due to these
effects.

\section*{Acknowledgments}
The authors are very grateful to D. Diakonov, A. Hosaka, S.i. Nam, and
M.V. Polyakov for helpful discussions and invaluable comments.  The
authors are also very thankful to Y. Azimov for the critical reading
of the present manuscript.  The present work is supported by the Korea
Research Foundation Grant funded by the Korean Government(MOEHRD)
(KRF-2006-312-C00507).  The work is also supported by the
Transregio-Sonderforschungsbereich Bonn-Bochum-Giessen, the
Verbundforschung (Hadrons and Nuclei) of the Federal Ministry for
Education and Research (BMBF) of Germany, the Graduiertenkolleg
Bochum-Dortmund, the COSY-project J\"ulich as well as the EU
Integrated Infrastructure Initiative Hadron Physics Project 
under contract number RII3-CT-2004-506078 and the DAAD.
\newpage

\begin{appendix}
\section{Electric transition densities\label{append:a}}
In the following, we list the explicit expressions of the densities
relevant in Eq.(\ref{model GE}) for each contribution to the electric-like
transition form factors:

\begin{eqnarray}
\mathcal{B}({\bm z}) & = & N_{c}\langle \mathrm{val}\mid{\bm z}
\rangle\langle{\bm z} \mid \mathrm{val}\rangle -
\frac{N_{c}}{2}\sum_{n} \mathrm{sign}(\varepsilon_{n})\langle
n\mid{\bm z}\rangle \langle{\bm z}\mid n\rangle, \\
\mathcal{C}({\bm z}) & = &
N_{c}\sum_{\varepsilon_{n}\neq\varepsilon_{\mathrm{val}}}\frac{1}{
\varepsilon_{n} - \varepsilon_{\mathrm{val}}} \langle \mathrm{val}
\mid{\bm z} \rangle \langle{\bm z} \mid n\rangle \langle n \mid
\gamma^{0} \mid \mathrm{val} \rangle \cr
&& +\; \frac{N_{c}}{2}\sum_{n,m}\mathcal{R}_{5}(\varepsilon_{n},
\varepsilon_{m})\langle  m\mid{\bm z}\rangle\langle{\bm z}\mid
n\rangle\langle n\mid\gamma^{0}\mid m\rangle,\\
\mathcal{I}_{1}({\bm z}) & = &
\frac{N_{c}}{6}\sum_{\varepsilon_{n}\neq\varepsilon_{\mathrm{val}}}
\frac{1}{\varepsilon_{n} -\varepsilon_{\mathrm{val}}} \langle
\mathrm{val}\mid{\bm z}\rangle{\bm \tau}\langle{\bm z}\mid n
\rangle\cdot \langle n\mid{\bm \tau}\mid \mathrm{val}\rangle \cr
&& + \; \frac{N_{c}}{12}\sum_{n,m}\mathcal{R}_{3}(\varepsilon_{n},
\varepsilon_{m})\langle m\mid{\bm z}\rangle{\bm \tau}\langle{\bm z}
\mid n\rangle\cdot\langle n\mid{\bm \tau}\mid m\rangle,\\
\mathcal{I}_{2}({\bm z}) & = & \frac{N_{c}}{4}
\sum_{\varepsilon_{n^{0}}}
\frac{1}{\varepsilon_{n^{0}} - \varepsilon_{\mathrm{val}}} \langle
\mathrm{val}\mid{\bm z} \rangle \langle{\bm z}\mid n^{0} \rangle
\langle n^{0}\mid \mathrm{val}\rangle \cr
&& + \; \frac{N_{c}}{4} \sum_{n,m^{0}}\mathcal{R}_{3}(\varepsilon_{n},
\varepsilon_{m^{0}}) \langle m^{0}\mid{\bm z}\rangle\langle{\bm z}
\mid n \rangle\langle n\mid m^{0}\rangle,\\
\mathcal{K}_{1}({\bm z}) & = & \frac{N_{c}}{6}
\sum_{\varepsilon_{n}\neq\varepsilon_{\mathrm{val}}}
\frac{1}{\varepsilon_{n}-\varepsilon_{\mathrm{val}}} \langle
\mathrm{val} \mid{\bm z}\rangle
{\bm \tau}\langle{\bm z}\mid n\rangle\cdot\langle
n\mid\gamma^{0}{\bm \tau}\mid \mathrm{val}\rangle\cr
 &  &\; +\frac{N_{c}}{12}\sum_{n,m}\mathcal{R}_{5}(\varepsilon_{n},
 \varepsilon_{m}) \langle m\mid{\bm z}\rangle{\bm \tau}\langle{\bm z}
 \mid n\rangle\cdot\langle n\mid\gamma^{0}{\bm \tau}\mid m\rangle,\\
\mathcal{K}_{2}({\bm z}) & = &
\frac{N_{c}}{4}\sum_{\varepsilon_{n^{0}}} \frac{1}{\varepsilon_{n^{0}}
  -\varepsilon_{\mathrm{val}}}\langle \mathrm{val}\mid{\bm z}\rangle
\langle{\bm z}\mid n^{0} \rangle\langle n^{0}\mid\gamma^{0}\mid
\mathrm{val} \rangle\cr
 &  & \;+ \frac{N_{c}}{4}\sum_{n,m^{0}}\mathcal{R}_{5}(\varepsilon_{n},
 \varepsilon_{m^{0}})\langle m^{0}\mid{\bm z}\rangle\langle{\bm z}\mid
 n\rangle\langle n\mid\gamma^{0}\mid
 m^{0}\rangle,
\end{eqnarray}
where the states $|\mathrm{val} \rangle$ and $|n\rangle$ stand for the
valence and sea quark states with the corresponding eigenenergies
$\varepsilon_{\textrm{val}}$ and $\varepsilon_{n}$ of the single-quark
Hamiltonian $h(U_{c})$ in Eq.(\ref{eq:diracham}), repecticvely.  The
summation $m^{0}$ runs over the vacuum states for $h(U=1)$ due
to the trivial embedding of Eq.(\ref{eq:embed}).  The regularization
functions $\mathcal{R}_{3}$ and $\mathcal{R}_{5}$ can be found in
Appendix~\ref{append:c}.
\section{Magnetic transition densities\label{append:b}}
In the following, we list the explicit expressions of the densities
relevant in Eq.(\ref{model GM}) for each contribution to the magnetic-like
transition form factors:
\begin{eqnarray}
\mathcal{Q}_{0}({\bm z}) & = & N_{c}\langle \mathrm{val} \mid{\bm z}
\rangle \gamma^{5}\{{\bm z}\times{\bm \sigma}\}\cdot{\bm \tau}
\langle{\bm z} \mid \mathrm{val}\rangle \cr
&& + \; N_{c}\sum_{n}\mathcal{R}_{1}(\varepsilon_{n}) \langle
n\mid{\bm z}\rangle\gamma^{5}\{{\bm z} \times{\bm  \sigma}\}\cdot{\bm
  \tau}\langle{\bm z}\mid n\rangle, \\
\mathcal{Q}_{1}({\bm z}) & = &
i\frac{N_{c}}{2}\sum_{\varepsilon_{n}\neq\varepsilon_{\mathrm{val}}}
\frac{ \textrm{sign}(\varepsilon_{n})}{
  \varepsilon_{n}-\varepsilon_{\mathrm{val}}} \langle n\mid{\bm z}
\rangle \gamma^{5}\{{\bm z} \times{\bm \sigma}\} \times{\bm \tau}
\langle \mathrm{val} \mid{\bm z} \rangle\cdot\langle \mathrm{val}
\mid{\bm \tau} \mid n\rangle\cr
 &  & + \; i\frac{N_{c}}{2}\sum_{n,m}\mathcal{R}_{4}(\varepsilon_{n},
 \varepsilon_{m}) \langle n\mid{\bm z}\rangle\gamma^{5}\{{\bm z}
 \times{\bm  \sigma}\}\times{\bm \tau}\langle{\bm z} \mid m \rangle
 \cdot \langle m\mid{\bm \tau}\mid n\rangle,\\
\mathcal{X}_{1}({\bm z}) & = &
N_{c}\sum_{\varepsilon_{n}\neq\varepsilon_{\mathrm{val}}}
\frac{1}{\varepsilon_{n} -\varepsilon_{\mathrm{val}}}\langle
\mathrm{val} \mid{\bm z}\rangle\gamma^{5}\{{\bm z}\times{\bm \sigma}\}
\langle{\bm z}\mid n\rangle\cdot\langle n\mid{\bm \tau}\mid
\mathrm{val}\rangle\cr
&  & +\; \frac{N_{c}}{2}\sum_{n,m}
\mathcal{R}_{5}(\varepsilon_{n},\varepsilon_{m}) \langle m\mid{\bm z}
\rangle \gamma^{5}\{{\bm z}\times{\bm \sigma}\}\langle{\bm z}\mid n
\rangle \cdot\langle n\mid{\bm \tau}\mid m\rangle,\\
\mathcal{X}_{2}({\bm z}) & = &
N_{c}\sum_{\varepsilon_{n^{0}}}\frac{1}{
  \varepsilon_{n^{0}}-\varepsilon_{\mathrm{val}}} \langle \mathrm{val}
\mid{\bm z}\rangle\gamma^{5}\{{\bm z}\times{\bm \sigma}\}\cdot{\bm
  \tau} \langle{\bm z}\mid n^{0}\rangle\langle n^{0}\mid  \mathrm{val}
\rangle\cr
 &  & + \;
 N_{c}\sum_{n^{0},m}\mathcal{R}_{5}(\varepsilon_{m},\varepsilon_{n^{0}})
 \langle m\mid{\bm z}\rangle\gamma^{5}\{{\bm z}\times{\bm \sigma}\}
 \cdot{\bm \tau} \langle{\bm z}\mid n^{0}\rangle\langle n^{0}\mid{\bm
   \tau} \mid m\rangle,\\
\mathcal{M}_{0}({\bm z}) & = & \frac{N_{c}}{3}
\sum_{\varepsilon_{n}\neq\varepsilon_{\mathrm{val}}}
\frac{1}{\varepsilon_{n} -\varepsilon_{\mathrm{val}}}\langle
\mathrm{val} \mid{\bm z}\rangle\gamma^{5}\{{\bm z}\times{\bm \sigma}\}
\cdot{\bm \tau}\langle{\bm z}\mid n\rangle\langle n\mid\gamma^{0}\mid
\mathrm{val}\rangle\cr
 &  & - \;\frac{N_{c}}{6}\sum_{n,m}\mathcal{R}_{2}(\varepsilon_{n},
 \varepsilon_{m}) \langle m\mid{\bm z}\rangle\gamma^{5}\{{\bm
   z}\times{\bm \sigma} \}\cdot{\bm \tau}\langle{\bm z}\mid n\rangle
 \langle n \mid\gamma^{0}\mid m\rangle,\\
\mathcal{M}_{1}({\bm z}) & = & \frac{N_{c}}{3}\sum_{\varepsilon_{n}
  \neq
  \varepsilon_{\mathrm{val}}}\frac{1}{\varepsilon_{n}-
  \varepsilon_{\mathrm{val}} }\langle \mathrm{val}\mid{\bm z} \rangle
\gamma^{5} \{{\bm z}\times{\bm \sigma}\}\langle{\bm z}\mid n\rangle
\cdot \langle n\mid\gamma^{0}{\bm \tau}\mid \mathrm{val}\rangle\cr
 &  & -\;\frac{N_{c}}{6}\sum_{n,m}\mathcal{R}_{2}(\varepsilon_{n},
 \varepsilon_{m}) \langle m\mid{\bm z}\rangle\gamma^{5}\{{\bm z}
 \times{\bm \sigma} \}\langle{\bm z}\mid n\rangle\cdot \langle n
 \mid\gamma^{0}{ \bm \tau}\mid m\rangle,\\
\mathcal{M}_{2}({\bm z}) & = &
\frac{N_{c}}{3}\sum_{\varepsilon_{n^{0}}} \frac{1}{\varepsilon_{n^{0}}
  -\varepsilon_{\mathrm{val}}}\langle \mathrm{val}\mid{\bm z}\rangle
\gamma^{5} \{{\bm z}\times{\bm \sigma}\}\cdot{\bm \tau} \langle{\bm z}
\mid n^{0}\rangle\langle n^{0}\mid\gamma^{0}\mid
\mathrm{val}\rangle\cr
 &  & - \;
 \frac{N_{c}}{3}\sum_{n,m^{0}}\mathcal{R}_{2}(\varepsilon_{n},
 \varepsilon_{m^{0}}) \langle n\mid{\bm z}\rangle\gamma^{5}\{{\bm z}
 \times{\bm \sigma}\} \cdot{\bm \tau}\langle{\bm z}\mid m^{0} \rangle
 \langle m^{0} \mid\gamma^{0}\mid n\rangle.
\end{eqnarray}
The regularization functions can be found in Appendix~\ref{append:c}.
\section{Regularization functions\label{append:c}}
The regularization functions in the electromagnetic transition densities
are given as follows:
\begin{eqnarray}
\mathcal{R}_{1}(\varepsilon_{n}) & = & -\frac{1}{2\sqrt{\pi}}
\varepsilon_{n} \int_{1/\Lambda^{2}}^{\infty}\frac{du}{
\sqrt{u}}e^{-u\varepsilon_{n}^{2}},\\
\mathcal{R}_{2}(\varepsilon_{n},\varepsilon_{m}) & = &
\int_{1/\Lambda^{2}}^{\infty}du\frac{1}{2\sqrt{\pi u}}\frac{
\varepsilon_{m}e^{-u\varepsilon_{m}^{2}}-\varepsilon_{n}e^{-u\varepsilon_{n}^{2}}}{
\varepsilon_{n}-\varepsilon_{m}},\\
\mathcal{R}_{3}(\varepsilon_{n},\varepsilon_{m}) & = & \frac{1}{
2\sqrt{\pi}}\int_{1/\Lambda^{2}}^{\infty}\frac{du}{\sqrt{u}}\left[
\frac{1}{u}\frac{e^{-\varepsilon_{n}^{2}u}-e^{-\varepsilon_{m}^{2}u}}{
\varepsilon_{m}^{2}-\varepsilon_{n}^{2}}-\frac{\varepsilon_{n}
e^{-u\varepsilon_{n}^{2}}+\varepsilon_{m}e^{-u\varepsilon_{m}^{2}}}{
\varepsilon_{m}+\varepsilon_{n}}\right],\\
\mathcal{R}_{4}(\varepsilon_{n},\varepsilon_{m}) & = &
\frac{1}{2\pi}\int_{1/\Lambda^{2}}^{\infty}du\int_{0}^{1}d\alpha\,
e^{-\varepsilon_{n}^{2}u(1-\alpha)-\alpha\varepsilon_{m}^{2}u}\frac{
\varepsilon_{n}(1-\alpha)-\alpha\varepsilon_{m}}{
\sqrt{\alpha(1-\alpha)}},\\
\mathcal{R}_{5}(\varepsilon_{n},\varepsilon_{m}) & = &
\frac{1}{2}\frac{\textrm{sign}\varepsilon_{n}-\textrm{sign}
\varepsilon_{m}}{\varepsilon_{n}-\varepsilon_{m}},\\
\mathcal{R}_{6}(\varepsilon_{n},\varepsilon_{m}) & = &
\frac{1-\textrm{sign}(\varepsilon_{n})
\textrm{sign}(\varepsilon_{m})}{\varepsilon_{n}-\varepsilon_{m}}.
\end{eqnarray}
\end{appendix}

\end{document}